%% file: main.tex
\documentclass[%
 reprint,
 amsmath,amssymb,
 aps,
]{revtex4-2}

\usepackage{graphicx}
\usepackage{dcolumn}
\usepackage{bm}
\usepackage{setspace}
\usepackage{amsmath}
\usepackage{mathtools}
\usepackage{xparse}
\usepackage{amssymb}
\usepackage{physics}
\usepackage{siunitx}
\usepackage{gensymb}
\usepackage{float}
\usepackage{mathrsfs}

\usepackage{xcolor}

\begin{document}

\preprint{APS/123-QED}

\title{Simulating the Galactic population of axion clouds around stellar-origin black holes: \\ Gravitational wave signals in the 10 -- 100 kHz band}

\author{Jacob R. Sprague$^{1}$, Shane L. Larson$^{1}$, Zhiyuan Wang$^{2}$, Shelby Klomp$^{2}$, Andrew Laeuger$^{3}$, George Winstone$^{2}$, Nancy Aggarwal$^{4}$, Andrew A. Geraci$^{2}$, and Vicky Kalogera$^{1}$\\ (The LSD Collaboration)}
\email{Contact author: JacobSprague2021@u.northwestern.edu}
\affiliation{$^{1}$Center for Interdisciplinary Exploration and Research in Astrophysics (CIERA), Department of Physics and Astronomy, Northwestern University, Evanston, Illinois 60208}
\affiliation{$^{2}$Center for Fundamental Physics, Department of Physics and Astronomy, Northwestern University, Evanston, Illinois 60208, USA}
\affiliation{$^{3}$The Division of Physics, Mathematics and Astronomy, California Institute of Technology, Pasadena, California 91125, USA}
\affiliation{$^{4}$Department of Physics and Astronomy, University of California, Davis, Davis, California 95616, USA}

\date{\today}

\input{Sections/Abstract.tex}

\maketitle

\input{Sections/Introduction.tex}
\input{Sections/AxionClouds.tex}
\input{Sections/GW.tex}
\input{Sections/BHPopulation.tex}
\input{Sections/Procedure.tex}
\input{Sections/Results.tex}
\input{Sections/Conclusion.tex}

\begin{acknowledgments}
We would like to thank Vedant Dhruv for making public his $Mathematica$ notebook for scalar bound-states in Kerr. We also thank Timothy Kovachy for clarifying issues of numerical precision when using $Mathematica$'s root-finding routines; and Richard Brito for several clarifying discussions on axion clouds \& their gravitational-wave emission. JS, AG, and SL are supported by the W.M. Keck Foundation. AG, GW, and NA are supported in part by NSF grants PHY-2110524 and PHY-2111544, the Heising-Simons Foundation, the John Templeton Foundation, and ONR Grant N00014-18-1-2370. NA is partially supported by the CIERA Postdoctoral Fellowship from the Center for Interdisciplinary Exploration and Research in Astrophysics at Northwestern University and the University of California-Davis.
SL is also supported by EPSRC International Quantum Technologies Network Grant EP/W02683X/1 and is grateful for EPSRC support through Standard Research Studentship (DTP) EP/R51312X/1. VK is supported by a CIFAR Senior Fellowship and through Northwestern University through the D.I. Linzer Distinguished University Professorship. A.L. is supported by the Fannie and John Hertz Foundation.
This work used 
the Quest computing facility at Northwestern. 
\end{acknowledgments}

\input{Sections/Appendix.tex}

\clearpage
\bibliography{bibliography}

\end{document}

%% file: Sections/Abstract.tex
\begin{abstract}
Ultralight scalar fields can experience runaway `superradiant' amplification near spinning black holes, resulting in a macroscopic `axion cloud' which slowly dissipates via continuous monochromatic gravitational waves. For a particular range of boson masses, $\mathcal{O}(10^{-11}$ -- $10^{-10})$ eV, an axion cloud will radiate in the $10$ -- $100$ kHz band of the Levitated Sensor Detector (LSD). Using fiducial models of the mass, spin, and age distributions of stellar-origin black holes, we simulate the present-day Milky Way population of these hypothetical objects. As a first step towards assessing the LSD's sensitivity to the resultant ensemble of gravitational wave signals, we compute the corresponding signal-to-noise ratios which build up over a nominal integration time of $10^{7}$ s, assuming the projected sensitivity of the $1$-m LSD prototype currently under construction, as well as for future $10$-m and $100$-m concepts. For a $100$-m cryogenic instrument, hundreds of resolvable signals could be expected if the boson mass $\mu$ is around $3\times10^{-11}$ eV, and this number diminishes with increasing $\mu$ up to $\approx 5.5\times10^{-11}$ eV. The much larger population of unresolved sources will produce a confusion foreground which could be detectable by a $10$-m instrument if $\mu \in (3-4.5)\times10^{-11}$ eV, or by a $100$-m instrument if $\mu \in (3-6)\times10^{-11}$ eV.

\end{abstract}

%% file: Sections/Introduction.tex
\section{\label{sec:level1}Introduction}


The era of gravitational-wave (GW) astronomy is in full-swing. During their first three observing runs, the GW interferometers Advanced LIGO and Advanced Virgo detected $90$ compact binary coalescences (CBC) involving neutron stars (NS) and stellar-mass black holes (BH) \cite{PhysRevX.9.031040}\cite{PhysRevX.11.021053}\cite{2021arXiv211103606T}$.$ The most notable events included the first NS-NS merger (GW170817) \cite{2017PhRvL.119p1101A}, the first highly-asymmetric binary (GW190412) \cite{2020PhRvD.102d3015A}, the first merger with an intermediate-mass BH remnant (GW190521) \cite{2020PhRvL.125j1102A}, and the first object in the mass gap separating the most massive neutron stars from the lowest-mass BH's (GW190814) \cite{2020ApJ...896L..44A}$.$ The first half of the fourth observing run has already seen a new lower-mass-gap event (GW230529) \cite{2024arXiv240404248T}.  \\ \\
\indent Adding to the excitement, evidence for a stochastic background has been reported in the $15$-year dataset from the North American Nanohertz Observatory for Gravitational Waves (NANOGrav) \cite{Agazie_2023}$.$ The most well-motivated scenario for the origin of this background is the extragalactic population of inspiralling supermassive BH binaries. \\ \\
\indent Finally, the launch of the Laser Interferometer Space Antenna (LISA) in the mid-$2030$'s will open up the millihertz band for exploration. The Galactic population of compact binaries, and the extragalactic population of supermassive BH binaries and extreme-mass-ratio-inspirals (EMRI's), are all highly anticipated LISA sources \cite{2023LRR....26....2A}. \\ \\
\indent These observatories cover multiple windows in the GW spectrum from the nanohertz up to several hundred Hz. The push to higher frequencies is now underway, with cosmic strings, axion clouds, primordial black hole (PBH) binaries, and early-universe stochastic backgrounds as the main science drivers \cite{2021LRR....24....4A}.   \\ \\
\indent One such concept, currently in development at Northwestern University, is the Levitated Sensor Detector (LSD). With sensitivity to GW's at tens to hundreds of kHz, the LSD employs optically-trapped micron-scale disks as GW sensors. The instrument is a Michelson interferometer with two perpendicular $1$-meter Fabry-P\'{e}rot arm cavities. In each arm, a disk is levitated at an antinode of a standing-wave formed by two counter-propagating beams. The trapped object behaves like a driven damped harmonic oscillator, with the corresponding trap frequency being widely-tunable with laser intensity. The periodic changes in arm-length induced by a GW manifest as a periodic shift in the position of the antinode. If the trap frequency matches the GW frequency, the levitated sensor is resonantly driven \cite{2013PhRvL.110g1105A} \cite{2022PhRvL.128k1101A} \cite{2022PhRvL.129e3604W}. \\ \\
\indent As a resonant detector, the LSD is well-suited to search for continuous monochromatic signals. A popular scenario involves the interaction between spinning black holes and `ultralight' bosonic fields -- i.e$.$ those with masses several orders-of-magnitude smaller than an electron-volt (eV). Such fields can extract rotational energy from spinning BH's via `superradiant amplification' of certain bound-states \cite{2007PhRvD..76h4001D}. The result is a macroscopic cloud of bosons all living in the same state -- commonly known as a  `gravitational atom' or `axion cloud' \cite{2019JCAP...12..006B}. These oscillating non-axisymmetric clouds generate continuous monochromatic GW's at a frequency primarily determined by the boson's mass. Tens to hundreds of kHz corresponds to $\mu = \mathcal{O}(10^{-11} - 10^{-10})$ eV.  \\ \\
\indent This scenario can be realized with physics beyond-the-Standard-Model (BSM). For example, a large number of ultralight fields may occur as a result of the compactification of extra dimensions \cite{2010PhRvD..81l3530A} \cite{2011PhRvD..83d4026A}. One of these may be the QCD axion -- the pseudoscalar boson proposed to solve the strong-CP problem \cite{PhysRevLett.38.1440}\cite{PhysRevLett.40.223}\cite{PhysRevLett.40.279}. The axion is a Goldstone boson of a spontaneously-broken global symmetry which acquires a small mass through non-perturbative effects. Its mass, $\mu$, is determined by the energy scale $f_{a}$ associated with the broken symmetry \cite{2011PhRvD..83d4026A},
\begin{equation}
    \label{eqn:PQmass}
    \mu \approx 6\times 10^{-10}\ \text{eV}\bigg(\frac{10^{16}\ \text{GeV}}{f_{a}} \bigg)
\end{equation}
where $10^{16}\ \text{GeV} \equiv \Lambda_{\text{GUT}}$ is the grand unification (GUT) scale. An axion of mass $\mathcal{O}(10^{-10})$ eV corresponds to $f_{a}$ being at the GUT scale. However, as we will see in Sec$.$ \ref{subsec:populations}, signals in the LSD band are only expected up to $\approx 32$ kHz, corresponding to a $6.6\times10^{-11}$ eV boson. \\ \\
\indent At boson masses $\mathcal{O}(10^{-11})$ eV, superradiance occurs optimally for BH's with masses between $0.1$ and a few solar masses. Sub-solar BH's may exist as PBH's \cite{2021RPPh...84k6902C}, and BH's in the $1 - 4\ M_{\odot}$ range might be formed dynamically in binary neutron star mergers \cite{2024Sci...383..275B}, accretion-induced collapsing neutron stars \cite{2023ApJ...954..212S}, or supernovae with unusually high fallback \cite{2016ApJ...821...38S} \cite{2020ApJ...890...51E}. \\ \\
\indent The $1 - 4\ M_{\odot}$ range of BH masses is gradually being populated by microlensing candidates \cite{2022A&A...662A..59K} and GW events such as GW190814 \cite{2020ApJ...896L..44A} and GW230529 \cite{2024arXiv240404248T}. Since the mass distribution for these objects is still unknown, we will limit our attention to stellar-origin BH's with masses between $5$ and $20\ M_{\odot}$, typical of BH's in X-ray binary systems.  \\ \\ 
\indent As a first step towards building the LSD search pipeline, we simulate the Galactic population of axion clouds with $5-20\ M_{\odot}$ BH hosts. The essential data returned by these simulations are the gravitational-wave frequency \& dimensionless strain amplitude emitted by each cloud. Together with the LSD's projected sensitivity curve, we estimate the number of resolvable signals, i.e$.$ those whose signal-to-noise ratio (SNR) rises above a given threshold after a coherent observation time $T_{\text{coh}} = 10^{7}$ s (a little less than four months). We adopt the idealization of a `freely-floating' detector orbiting the Milky Way at the same radius as the Solar System, but not situated on a rotating planet orbiting a star. In doing so, we neglect the amplitude and frequency modulations induced by the Earth's sidereal rotation and by its orbital motion in the Solar System. Our results establish a baseline from which a more in-depth analysis, including the aforementioned modulations, can be undertaken in future work.  \\ \\
\indent In Sections II \& III, we introduce the essential physics of axion clouds and their GW emission. To simulate the population of axion clouds, we require a model of the stellar-origin BH population. The parameters of a black hole -- mass, spin, age, and location in the Milky Way -- are taken to be independent random variables, and we discuss their distributions in Section IV. The procedure for determining whether a BH of given mass, spin, and age presently hosts an axion cloud is described in Section V. The simulated cloud populations, and the corresponding ensembles of GW signals, are discussed in Section VI. Section VII provides a summary of the results, as well as tasks for future work. Throughout the paper, we adopt the metric signature $(-, +, +, +)$, and we retain all factors of $G$, $c$, and $\hbar$. We hope our decision not to set physical constants to unity will make this work more accessible to those unaccustomed to the conventions of fundamental physics theory.

%% file: Sections/AxionClouds.tex
\section{\label{sec:superradiance}Superradiant bound-states}
\begin{figure*}
    \centering
    \includegraphics[width=0.6\textwidth]{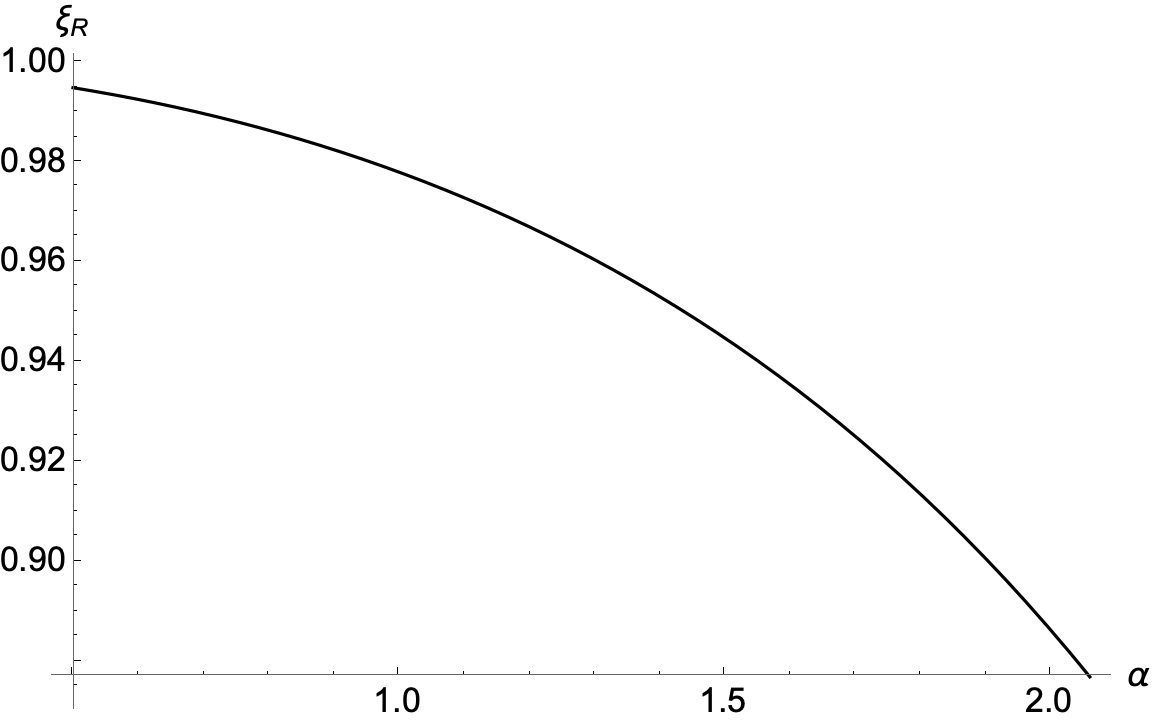}
    \caption{The real part $\xi_{R}$ of the $n = 5$ bound-state, for BH spin $\chi = 0.995$, plotted up to the associated maximum superradiant value of the coupling parameter, $\alpha_{\text{max}} = 2.06$.} 
    \label{fig:xir1}
\end{figure*}

\begin{figure*}
    \centering
    \includegraphics[width=0.7\textwidth]{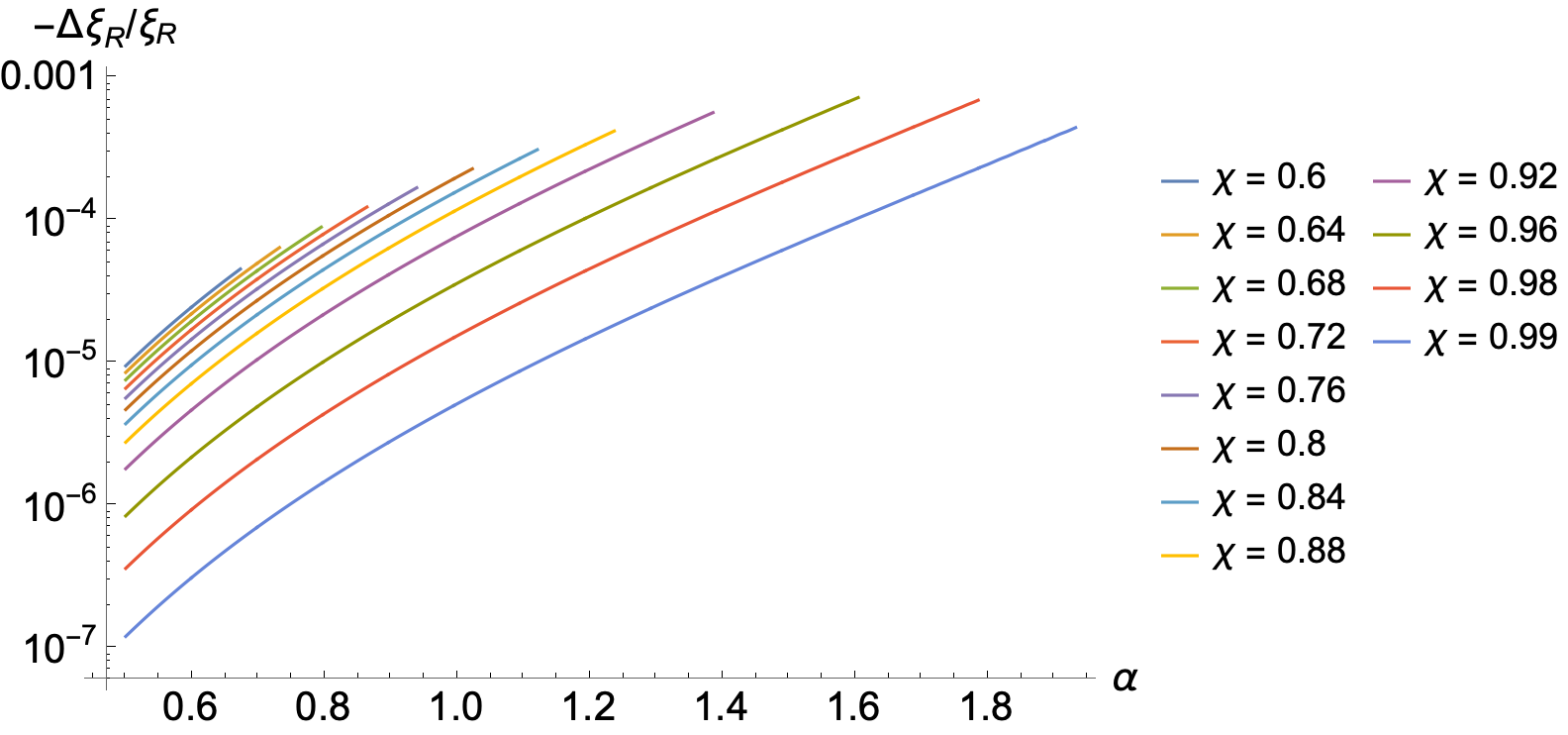}
    \caption{The $\xi_{R}$ curve in Fig$.$ \ref{fig:xir1} corresponds to a BH spin of $\chi = 0.995$. As mentioned in the main text, there are small spin-dependent corrections to $\xi_{R}$. Had we plotted several curves in Fig$.$ \ref{fig:xir1}, each corresponding to a different spin, they would lie so close together as to be almost indistinguishable. To spread them out, we plot the fractional deviations between the curve in Fig$.$ \ref{fig:xir1} and the corresponding curves for a variety of spins. As indicated on the y-axis, the fractional deviations are negative: For a given $\alpha$, lower BH spin implies lower $\xi_{R}$; In turn, this implies a lower GW frequency via Eq$.$ \ref{eqn:gwfreq}.} 
    \label{fig:xir2}
\end{figure*}

The creation of macroscopic clouds around spinning black holes can occur for any massive bosonic field. The simplest scenario, and the one we adopt, is that of an electrically-neutral massive scalar field freely propagating in the Kerr spacetime; We denote the BH mass and dimensionless spin by $M$ and $\chi \equiv Jc/(GM^{2})$, respectively ($J$ is the BH angular momentum). We also assume no self-interactions to avoid complications such as the bosenova instability \cite{2011PhRvD..83d4026A}. The scalar field then obeys the Klein-Gordon equation \cite{2007PhRvD..76h4001D},
\begin{equation}
    \Big[g^{\mu \nu}\nabla_{\mu}\nabla_{\nu} - m_{*}^{2} \Big]\Phi(\Vec{x}, t) = 0
    \label{kleingordon}
\end{equation}
where the constant $m_{*}$ has dimensions of inverse length; In the quantum theory of a scalar field, the physical meaning of $m_{*}$ is $1/\lambda_{c}$, where $\lambda_{c} \equiv \hbar/(mc) = \hbar c/\mu$ is the reduced Compton wavelength of the boson, $m$ is the mass of the particle, and $\mu = mc^{2}$. \\ \\
\indent In Boyer-Lindquist coordinates, and with the ansatz
\begin{equation}
    \Phi(\Vec{x}, t) = e^{-i \omega t}e^{i m \phi} S(\theta) R(r)
\label{kgansatz}
\end{equation}
the Klein-Gordon equation separates into two ordinary differential equations (ODE's) for $R(r)$ and $S(\theta)$. We seek a bound-state solution which is `in-going' at the event horizon -- i.e. a solution which goes to zero at infinity and looks like an in-going-wave at the horizon. The in-going boundary condition causes the eigenfrequency $\omega$ to be complex,
\begin{equation}
    \label{complexomega}
    \omega = \omega_{R} + i\omega_{I}
\end{equation}
with the consequence that bound-states must either grow or decay:
$$e^{-i \omega t} = e^{-i (\omega_{R} + i\omega_{I}) t} = e^{-i \omega_{R} t}e^{\omega_{I} t}$$
$$\implies \Phi(\Vec{x}, t) = e^{\omega_{I} t}\big[e^{-i \omega_{R} t}e^{i m \phi} S(\theta) R(r)\big] $$
\indent For $\omega_{I} > 0$, the field amplitude grows exponentially. A necessary and sufficient condition for the growth of a bound-state with azimuthal number $m$ is that the event horizon's angular speed $\Omega_{\text{H}}$ (times $m$) be faster than the oscillation of the field \cite{2007PhRvD..76h4001D},
\begin{equation}
    \omega_{R} < m\Omega_{\text{H}}
    \label{basicsr}
\end{equation}
\indent This requirement is called the `superradiance condition'. As the field amplitude grows, the BH loses rotational energy, and $\Omega_{\text{H}}$ decreases until the inequality becomes an equality. At that point, the superradiant growth ceases, and the resultant bound-state slowly dissipates by emitting GW's. \\ \\
\indent It is conventional to define a dimensionless `coupling parameter' $\alpha$ as the ratio of the BH's gravitational radius $r_{g}$ to the reduced Compton wavelength $\lambda_{c}$ of the scalar field:
\begin{equation}
    \alpha \vcentcolon= \frac{r_{g}}{\lambda_{c}} = \frac{GM}{c^{2}} \frac{\mu}{\hbar c} = \frac{GM\mu}{\hbar c^{3}}
\end{equation}
\indent The `weak-coupling' limit, defined by $\alpha \ll 1$, corresponds to the Compton wavelength of the boson being much larger than the characteristic size $r_{g}$ of the BH. In this limit, the bound-state energy, given by the real part of $\omega$, can be written in closed-form \cite{2019PhRvD..99d4001B}:
\begin{eqnarray}
     \label{det}
    \hbar \omega_{R} = \mu \bigg[1 - \frac{\alpha^{2}}{2n^{2}} + \bigg(\frac{2l - 3n + 1}{l + 1/2} - \frac{1}{8}\bigg)\frac{\alpha^{4}}{n^{4}} + \nonumber \\ \frac{2m\chi\alpha^{5}}{n^{3}l(l + 1/2)(l + 1)} + ...\bigg]
\end{eqnarray}
\indent The small `fine-structure' corrections beyond the leading 1 depend on the angular momentum of the cloud and the spin of the BH. The quantity in large square brackets depends on the BH \& boson masses only through their dimensionless product $\alpha$. This motivates the introduction of a dimensionless eigenfrequency $\xi = \xi_{R} + i\xi_{I}$:
\begin{equation}
    \xi_{R} \equiv \frac{\hbar\omega_{R}}{\mu} \ \ \ \ \ \ \xi_{I} \equiv \frac{\hbar\omega_{I}}{\mu}
\end{equation}
\indent Once we have computed $\xi$ over a sufficiently large region of the $(\alpha, \chi)$ parameter space for all bound-states \{$n, l, m$\} of interest, we can freely plug-in any BH masses and axion masses of our choosing. For example, taking $M=10^{7}\ M_{\odot}$ and $\mu = 10^{-17}$ eV, we get $\alpha = 0.748$. The same value is obtained taking $M=10\ M_{\odot}$ and $\mu = 10^{-11}$ eV. The essential consequence is that, for a given BH spin $\chi$, the same set of superradiant bound-states exists for both scenarios. \\ \\
\indent From a practical point of view, this also means the superradiance condition
\begin{equation}
    \label{eqn:SRineqdim}
    \xi_{R} < \frac{m\chi}{2\alpha \big[1 + \sqrt{1 - \chi^{2}}\big]} \equiv \xi_{\text{crit}}
\end{equation}
becomes a tool for rapidly determining, for a given parameter set \{$\mu, M, \chi$\}, which states are superradiant. \\ \\
\indent Additionally, far from the BH where relativistic effects are negligible, the radial equation reduces to ($r$ measured in units of $\lambda_{c}$) \cite{2019JCAP...12..006B}
\begin{equation}
    \label{eqn:hydrogen}
    \Bigg[-\frac{1}{2r^{2}}\frac{d}{dr}\bigg(r^{2}\frac{d}{dr}\bigg) - \frac{\alpha}{r} + \frac{l(l + 1)}{2r^{2}} + \frac{1 - \xi^{2}}{2} \Bigg]R(r) = 0
\end{equation}
which is the radial Schr\"{o}dinger equation for a non-relativistic particle in a Coulomb potential $V(r) = \alpha/r$ -- hence the moniker `gravitational atom'. The $\alpha^{2}/(2n^{2})$ term in Eq$.$ \ref{det} is precisely the `hydrogen atom' solution to Eq$.$ \ref{eqn:hydrogen}. The complete small-$\alpha$ solutions to the full Klein-Gordon equation have been computed order-by-order using the method of matched asymptotic expansions \cite{2019JCAP...12..006B} (and we have provided an example of Eq$.$ \ref{eqn:xirNR} in Figs$.$ \ref{fig:xir1} \& \ref{fig:xir2}),
\begin{eqnarray}
    \label{eqn:xirNR}
    \xi_{R} = 1 - \frac{\alpha^{2}}{2n^{2}} + \bigg[\frac{2l - 3n + 1}{l + 1/2} - \frac{1}{8}\bigg]\frac{\alpha^{4}}{n^{4}} + \ \ \ \ \ \ \ \ \ \ \ \ \ \ \ \ \ \nonumber \\ \frac{2m\chi\alpha^{5}}{n^{3}l(l + 1/2)(l + 1)} + ... \ \ \ \ \ \ \ \ \ \ \ \ \ \ \ \  \ \ \ \ \ \ \ \ \ \ \  \ \ \ \
\end{eqnarray}
\begin{eqnarray}
    \label{eqn:xiiNR}
    \xi_{I} = 2\Big(1 + \sqrt{1 - \chi^{2}}\Big)\bigg[\frac{m\chi}{2\alpha \big(1 + \sqrt{1 - \chi^{2}}\big)} - \xi_{R} \bigg]\alpha^{4l + 5} \ \ \nonumber \\
    \times \frac{2^{4l + 1}(n + l)!}{n^{2l + 4}(n - l - 1)!}\bigg(\frac{l!}{(2l)!(2l + 1)} \bigg)^{2} \ \ \ \ \ \ \ \ \ \ \ \ \ \ \ \ \ \ \ \ \ \ \nonumber \\ \times \prod_{j = 1}^{l}\bigg[j^{2}(1 - \chi^{2}) + \Big[m\chi - 2\alpha\xi_{R}\Big(1 + \sqrt{1 - \chi^{2}}\Big)\Big]\bigg] \ \ \ \ \ \
\end{eqnarray}
\indent In general, $\xi$ is a function of \{$n, l, m, \alpha, \chi$\}. For fixed $\alpha$ and $\chi$, the superradiance rate is largest when $m = l = n - 1$. We consider only such bound states in our simulations of the Galactic axion cloud population, reducing the parameter list to \{$n, \chi, \alpha$\}. \\ \\
\indent Since our fiducial model of the Galactic BH population will assume $M \geq 5\ M_{\odot}$, as well as boson masses $\mathcal{O}(10^{-11} - 10^{-10})$ eV, the corresponding values of $\alpha$ are always greater than $1$, but still of order unity. In this `intermediate' regime, there are no closed-form solutions for $\xi$. As detailed in Appendix A, we must resort to the series-solution method for solving the radial Klein-Gordon equation. The coefficients of the infinite-series ansatz obey a three-term recurrence relation whose solution is equivalent to the solution of a corresponding non-linear continued-fraction equation \cite{2007PhRvD..76h4001D} \cite{1967SIAMR...9...24G}. \\ \\
\indent Denoting the peak mass of the cloud as $M_{c}$, the cloud's growth timescale is given by \cite{2020PhRvD.102f3020Z}
\begin{equation}
    \label{eqn:timescale}
    \tau_{c} \equiv \tau_{nlm}\ln N = \tau_{nlm}\ln\bigg(\frac{M_{c}c^{2}}{\mu}\bigg)
\end{equation}
with $N$ the number of bosons in the cloud, and $\tau_{nlm}$ the reciprocal of the superradiance rate,
\begin{equation}
    \tau_{nlm} \equiv \frac{1}{\Gamma_{nlm}}, \ \ \ \ \Gamma_{nlm} \equiv 2\omega_{I}
\end{equation}

\indent $\tau_{nlm}$ is the e-folding timescale, and we follow the authors of \cite{2020PhRvD.102f3020Z} in taking $\tau_{c}$ as the time to fully grow the bound-state. The factor of two in $\Gamma_{nlm}$ occurs because the cloud's density is proportional to the $00$-component of the stress-energy, $\rho \propto T^{0}_{0} \propto \exp(2\omega_{I}t)$. \\ \\
\indent As the cloud grows, the BH gradually loses mass and angular momentum. The growth timescales are long enough to permit an adiabatic treatment of the BH's evolution \cite{2019PhRvD..99h4042I}. The metric can be thought of as Kerr with slowly changing $M$ and $\chi$. Denoting the initial BH parameters as ($M_{i}, \chi_{i}$), the cloud's mass is 
\begin{equation}
    \label{eqn:cloudmass}
    M_{c} \equiv M_{i} - M_{f}
\end{equation}
and the hole's final mass $\&$ spin ($M_{f},\ \chi_{f}$) are given by \cite{2019PhRvD..99h4042I}\cite{2021PhRvD.104d4011Y}
\begin{equation}
    \label{eqn:BHfinalmass}
    M_{f} = M_{i}\Bigg[\frac{m^{3} -\sqrt{m^{6} - 16m^{2}\xi_{R}^{2}\alpha_{i}^{2}\big(m - \xi_{R}\alpha_{i}\chi_{i}\big)^{2}}}{8\xi_{R}^{2}\alpha_{i}^{2}(m - \xi_{R}\alpha_{i}\chi_{i})} \Bigg]
\end{equation}
\begin{equation}
    \label{eqn:BHfinalspin}
    \chi_{f} = \bigg(\frac{M_{i}}{M_{f}}\bigg)^{2}\Bigg(\chi_{i} - \frac{m M_{c}}{\xi_{R}\alpha_{i} M_{i}} \Bigg)
\end{equation}
\indent Since our simulation of the Galactic axion cloud population requires us to follow the evolution of each BH-cloud system, -- of which there could be millions --  we save computation time by relying on these expressions for the final BH parameters. \\ \\
\indent The final mass $\&$ spin become the new parameters ($M_{f} \rightarrow M_{i}$, $\chi_{f} \rightarrow \chi_{i}$) for determining which bound-state will grow after the present cloud has dissipated. For our simulations, the superradiance condition (Eq$.$ \ref{eqn:SRineqdim}) is used to determine, from the set $\{1, 2, 3, ...\}$, the smallest value of $m$ for which superradiance occurs. The final state of the BH-boson system at the cessation of cloud growth is determined by Eqs$.$ \ref{eqn:cloudmass}, \ref{eqn:BHfinalmass}, and \ref{eqn:BHfinalspin}.

%% file: Sections/GW.tex
\section{\label{sec:GWs}Gravitational waves from axion clouds}
\indent At a particle-physics level, GW production by axion clouds can be understood in terms of two processes: annihilation of two bosons to a single graviton (with the BH absorbing the recoil momentum), and downward-transitions between bound-states \cite{2015PhRvD..91h4011A} \cite{2014PTEP.2014d3E02Y}. However, just as superradiance is a purely classical kinematic effect, the GW emission can also be understood classically in terms of the cloud's time-dependent quadrupole moment. That being said, the GW signals considered in this work correspond to the annihilation channel. \\ \\
\indent Since our simulation of the Galactic axion cloud population requires us to compute the GW amplitude for each cloud, -- of which there could be millions --  we save computation time by relying on semi-analytic formulas for the amplitudes \cite{2021PhRvD.104d4011Y} \cite{2014PTEP.2014d3E02Y}. Following \cite{2019PhRvD..99h4042I}, the GW signal seen by a detector with perpendicular arms takes the general form
\begin{equation}
    h(t) = F_{+}(t)a_{+}\cos[\Phi(t)] + F_{\cross}(t)a_{\cross}\sin[\Phi(t)]
    \label{ht}
\end{equation}
where $F_{+}(t)$ and $F_{\cross}(t)$ are the detector's angular pattern functions, and the phase $\Phi(t)$ encodes the frequency evolution in the detector frame,
\begin{equation}
    \begin{split}
        \Phi(t) & = \Phi_{0} + 2\pi\int_{T_{0}}^{T}f(T')dT' \\
         & = \Phi_{0} + 2\pi\bigg[f_{0}\big(T - T_{0}\big) + \frac{1}{2}\Dot{f}_{0}\big(T - T_{0}\big)^{2} + \ldots \bigg]
    \end{split}
\end{equation}
where $\Dot{f}$ is the frequency derivative, quantities with a zero-subscript are their values at a reference time $T_{0}$, and $T(t)$ is the time at the Solar System barycenter, related to the time at the detector $t$ by the addition of the R\o mer, Shapiro, and Einstein delays. \\ \\ 
\indent The amplitudes $a_{+ / \cross}$ are expanded in terms of spheroidal harmonics with spin-weight $s = -2$,
\begin{equation}
    a_{+ / \cross} = -\sum_{\Tilde{l} \geq 2l} h_{0}^{(\Tilde{l})}\Big[\mathstrut_{-2}S_{\Tilde{l},\Tilde{m},\Tilde{\omega}} \pm \mathstrut_{-2}S_{\Tilde{l},-\Tilde{m},-\Tilde{\omega}} \Big] 
    \label{apm}
\end{equation} 
where $\Tilde{\omega} = 2\omega_{R}$ is the GW angular frequency, the parameters ($l, m$) refer to the scalar bound-state, and ($\Tilde{l}, \Tilde{m}$) refer to the GW modes, with $\Tilde{l} \geq 2l$ and $\Tilde{m} = 2m$. For each mode, there is a polarization-independent characteristic amplitude $h_{0}^{(\Tilde{l})}$ \cite{2019PhRvD..99h4042I}:
\begin{equation}
    \label{h0}
    h_{0}^{(\Tilde{l})} = \frac{c^4}{G}\frac{M_c}{M_{f}} \frac{1}{2 \pi^2 M_{f} f^2 d} \mathcal{A}_{\Tilde{l} \Tilde{m}}(\alpha_{i}, \chi_{i})
\end{equation}
where $f$ is the GW frequency, $d$ is the source distance, and the $\mathcal{A}_{\Tilde{l} \Tilde{m}}(\alpha, \chi)$ are dimensionless numerical factors which measure how much energy is carried by each mode. The corresponding luminosity in each mode is given by
\begin{equation}
    \label{edot1}
    \Dot{E}_{\text{GW}}(\Tilde{l}, \Tilde{m}, \Tilde{\omega}) = \frac{c^{5}}{4\pi G} \bigg(\frac{c^{3}}{GM_{f} \Tilde{\omega}} \bigg)^{2} \frac{M_{c}^{2}}{M_{f}^{2}} \mathcal{A}_{\Tilde{l}\Tilde{m}}^{2}(\alpha_{i}, \chi_{i})
\end{equation}
\indent In principle, the coefficients $\mathcal{A}_{\Tilde{l} \Tilde{m}}$ must be computed numerically by solving the Teukolsky equation governing linear perturbations of the Kerr metric. The authors of \cite{2021PhRvD.104d4011Y} express $\Dot{E}_{\text{GW}}$ in the form
\begin{equation}
    \label{edot2}
    \Dot{E}_{\text{GW}} = \frac{c^{5}}{G} \frac{M_{c}^{2}}{M_{f}^{2}} \frac{d \Tilde{E}}{dt}
\end{equation}
and invoke an analytic solution for $d\Tilde{E}/dt$ which is formally valid for $\alpha \ll l$, and which remains a good approximation up to $\alpha \sim l$ \cite{2014PTEP.2014d3E02Y}:
\begin{equation}
    \label{eqn:dedt}
    \frac{d\Tilde{E}}{dt} = \frac{16^{l + 1} l(2l - 1) \Gamma^{2}(2l - 1) \Gamma^{2}(n + l + 1)\alpha_{f}^{4l + 10}}{n^{4l + 8}(l + 1)\Gamma^{4}(l + 1)\Gamma(4l + 3) \Gamma^{2}(n - l)}
\end{equation}
where $\Gamma$ is the gamma function, and $\alpha_{f}$ denotes the value of $\alpha$ corresponding to the final mass of the BH (i.e$.$ after the cloud has finished growing),
\begin{equation}
    \alpha_{f} = \alpha_{i}\frac{M_{f}}{M_{i}}
\end{equation}
\indent Comparing Eqs$.$ \ref{edot1} and \ref{edot2}, we see that $\mathcal{A}_{\Tilde{l}\Tilde{m}} \propto \sqrt{d\Tilde{E}/dt}$, allowing us to express $h_{0}^{(\Tilde{l})}$ directly in terms of $d\Tilde{E}/dt$. Restricting ourselves to the dominant mode $\Tilde{m} = \Tilde{l} = 2l$, we obtain a closed-form solution for the characteristic amplitude, which we use without abandon to compute the GW amplitudes of the axion clouds resulting from our simulations (we will drop the superscript $(2l)$ henceforth),
\begin{equation}
    \label{eqn:h0}
    h_{0}^{(2l)}(d) = \frac{GM_{c}}{c^{2}d} \frac{2\sqrt{\pi}M_{i}}{\xi_{R}\alpha_{i}M_{f}} \sqrt{\frac{d \Tilde{E}}{dt}}
\end{equation}
\indent The corresponding GW frequency is given by
$$f = \frac{\Tilde{\omega}}{2\pi} = \frac{1}{2\pi} \frac{2\mu}{\hbar}\xi_{R} \equiv f_{0}\xi_{R} $$
\begin{equation}
    f = f_{0}\xi_{R}
    \label{eqn:gwfreq}
\end{equation}
where we've introduced the zeroth-order frequency $f_{0} = \omega_{0}/2\pi$, $\omega_{0} \equiv 2\mu/\hbar$. \\ \\
\indent It is often remarked that the GW frequency is proportional to twice the axion mass, $f \propto 2\mu$. We see that this is, indeed, true in the small-$\alpha$ limit by noting that $\xi_{R} \rightarrow 1$ as $\alpha \rightarrow 0$ (Fig$.$ \ref{fig:xir1}, Eq$.$ \ref{eqn:xirNR}). The frequency monotonically decreases with increasing $\alpha$, and for axion clouds in the kHz band, with stellar-mass BH hosts (where $\alpha$ is generically greater than 1), GW frequencies can be upwards of $10\%$ smaller than the nominal value $f_{0}$. \\ \\
\indent Eq$.$ \ref{eqn:gwfreq} gives the frequency as measured in the rest-frame of the axion cloud. For an observer located elsewhere in the Milky Way, the measured signal is Doppler-shifted due to the differential rotation of the Galaxy. We assume all bodies in the Galaxy move in the azimuthal direction, $\Vec{v} = v_{\phi}\hat{\phi}$, and we assume the following Galactic rotation curve \cite{2013A&A...549A.137I} ($r$, in kpc, is the cylindrical radial distance from the Galactic center):
\begin{equation}
    \label{rotationcurve}
    v_{\phi}(r) \  (\text{km/s}) = \begin{cases}
        265 - 1875(r - 0.2)^{2} \ \ \ \ \ r < 0.2 \\
        225 + 15.625(r - 1.8)^{2} \ \ \ 0.2 < r < 1.8 \\
        225 + 3.75(r - 1.8) \ \ \ \ \ \ \ 1.8 < r < 5.8 \\
        240 \ \ \ \ \ \ \ \ \ \ \ \ \ \ \ \ \ \ \ \ \ \ \ \ \ \ \ \ r > 5.8
    \end{cases}
\end{equation}
\indent Denoting the source-frame frequency as $f_{s}$, the non-relativistic Doppler-shifted frequency we observe is
\begin{equation}
    \label{eqn:doppler}
    f_{\text{obs}} = \bigg(1 - \frac{v_{r}}{c} \bigg)f_{s}
\end{equation}
where $v_{r}$ is the line-of-sight component of the relative velocity between source and observer. $v_{r}$ is defined to be positive when the source and observer are moving away from each other.  \\ \\
\indent When a cloud finishes growing, it emits GW's whose initial amplitude $h_{0}$ is given by Eq$.$ \ref{h0}. As the cloud dissipates, the amplitude decreases as \cite{2020PhRvD.102f3020Z}
\begin{equation}
    \label{eqn:h_of_t}
    h(t) = \frac{h_{0}}{1 + t/\tau_{\text{GW}}}
\end{equation}
where $\tau_{\text{GW}}$ is the time for $h$ to drop to half its initial value.

%% file: Sections/BHPopulation.tex
\section{\label{sec:galaxy} The Galactic population of isolated stellar-origin black holes}

With the results of the previous sections in hand, we can follow the `superradiance history' of any given BH -- i.e. we can determine the sequence of scalar field bound-states, their growth \& dissipation timescales, the BH mass and spin decrements, and, above all, the GW frequency \& amplitude of each successive cloud. To simulate the entire Galactic population of axion clouds, we must assign each BH a mass, spin, age, and location -- taken to be independent random variables -- in accordance with known or assumed distributions.  \\ \\
\indent Our knowledge of the stellar-origin BH mass distribution is informed by mass measurements in X-ray binary systems \cite{2011ApJ...741..103F} \cite{MILLER20151} \cite{2020ApJ...898..143S}, microlensing events \cite{2022ApJ...933...83S}, and astrometry \cite{2023MNRAS.518.1057E} \cite{2023AJ....166....6C}, as well as through modelling of the complex physics of core-collapse supernovae \cite{2012ApJ...749...91F}. Known BH's typically have masses between $5\ M_{\odot}$ and $20\ M_{\odot}$, and power-law models are favored when fitting the mass function of low-mass X-ray binaries \cite{2011ApJ...741..103F}. Not coincidentally, the massive stars which produce BH remnants are also characterized by a power-law distribution, $\psi(M)dM \propto M^{-2.35}dM$ -- the `Salpeter' function. We will assume $M_{\text{BH}}$ to be Salpeter-distributed on the interval $5 - 20\ M_{\odot}$.  \\ \\
\indent BH spins have been measured in several X-ray binaries \cite{2021ARA&A..59..117R}, but none have been measured for isolated BH's. In the case of binaries, the distribution of spin magnitudes is more-or-less uniform, so we take the BH spin to be uniformly distributed, $\chi \sim U[0, 1]$. \\ \\
\indent The stellar content of the Milky Way can be divided into three primary regions -- the thin disk, the thick disk, and the central bulge. The age distribution of stellar-origin BH's is tied to the star formation history in each region. As the Milky Way's star formation history is a topic of ongoing research, we take an agnostic approach by assigning each BH an age of $10^{x}$ yr, with $x$ uniformly distributed on an interval which varies between the three Galactic regions. For the thin disk and thick disk, we take $x \sim U[3,\ \log_{10}(8\cross10^{9})]$ and $x \sim U[3, 10]$, respectively \cite{2017ApJ...837..162K}. For the bulge, we assign each BH an age $10^{x}$ yr, with $x \sim U[9,\ \log_{10}(13\cross10^{9})]$ \cite{doi:10.1146/annurev-astro-032620-021917}.  \\ \\ 
\indent We assume black holes are distributed in space according to the mass profiles of the disks \& bulge described in Ref$.$ \cite{2011MNRAS.414.2446M}. Both disks have the same axisymmetric form, with the corresponding scale lengths, scale heights, and surface densities quoted in Table \ref{table}:
\begin{equation}
    \rho_{\text{disk}}(r, z, \phi) = \frac{\Sigma_{\text{d, 0}}}{2 z_{\text{d}}} e^{-\abs{z}/z_{\text{d}}} e^{-r/R_{\text{d}}}
    \label{rhodisks}
\end{equation}
\indent The bulge is also axisymmetric, with the corresponding parameters also given in Table \ref{table}:
\begin{equation}
    \label{rhobulge}
    \rho_{\text{b}} = \frac{\rho_{\text{b,0}}}{\Big(1 + \frac{r'}{r_{0}}\Big)^{\alpha}}e^{-(r'/r_{\text{cut}})^{2}}
\end{equation}
$$r' \equiv \sqrt{r^{2} + (z/q)^{2}}$$
\begin{table}[H]
\centering
\begin{tabular}{ |p{4.0cm}||p{2.5cm}| }
\hline
Disk Parameters & Value \\
\hline
*$R_{\text{d, thin}}$ (kpc)  & 3.00  \\
*$R_{\text{d, thick}}$ (kpc) &   3.29 \\
*$\Sigma_{\text{d,0, thin}}$ ($M_{\odot}\ \text{pc}^{-2}$) & 741  \\
*$\Sigma_{\text{d,0, thick}}$ ($M_{\odot}\ \text{pc}^{-2}$) & 238  \\
$z_{\text{d, thin}}$ (kpc) & 0.3 \\
$z_{\text{d, thick}}$ (kpc)   &0.9 \\
*Solar radius $R_{\odot}$ (kpc) & 8.29 \\
\hline
Bulge Parameters & Value \\
\hline
*$\rho_{\text{b,0}}$ ($M_{\odot}\ \text{pc}^{-3}$) & 95.5 \\
$\alpha$ & 1.8 \\
$r_{0}$ (kpc) & 0.075 \\
$r_{\text{cut}}$ (kpc) & 2.1 \\
$q$ & 0.5 \\
\hline
\end{tabular}
\caption{Physical parameters for the empirical stellar-mass distribution of the Milky Way inferred in Ref$.$ \cite{2011MNRAS.414.2446M}. The values for those with an asterisk (*) are the means marginalized over all other parameters in the model. The disk scale heights are the best-fitting values from Ref$.$ \cite{2008ApJ...673..864J}, and the stellar bulge model is an axisymmetric modification of the result from Ref$.$ \cite{2002MNRAS.330..591B} in which the assumption of a constant mass-to-light ratio in the bulge was used to convert photometric data into a mass model.}
\label{table}
\end{table}
\indent We apportion the BH's among the three Galactic regions according to the fractions $f_{\text{thin}}$, $f_{\text{thick}}$, and $f_{\text{bulge}}$, defined by $f_{i} = M_{i}/\sum_{i}M_{i}$, $i \in $ \{thin, thick, bulge\}. The disk masses are obtained by integrating $\rho_{\text{disk}}$, with the radial integral cut-off at $25$ kpc, and the vertical integral cut-off at $3$ scale heights. This gives $3.97\cross10^{10}\ M_{\odot}$ and $1.5\cross10^{10}\ M_{\odot}$ for the thin and thick disks, respectively. We take the bulge mass to be $8.9\cross10^{9}\ M_{\odot}$, the value quoted in \cite{2011MNRAS.414.2446M}. The corresponding $f_{i}$ are $62\%$, $24\%$, $14\%$, respectively. We will assume the Galactic population of $N_{\text{BH}}$ BH's to be apportioned likewise: $62\%$ in the thin disk, $24\%$ in the thick disk, and $14\%$ in the bulge.

%% file: Sections/Procedure.tex
\section{\label{sec:procedure} Simulation procedure}
The simulation is a procedure by which, for a given axion mass, and from an initial population of $N_{\text{BH}}$ BH's sprinkled throughout the Milky Way, we determine the number $N_{c}$ of extant axion clouds. Each simulation outputs the physical properties, distances, and the GW frequencies \& amplitudes of the $N_{c}$ clouds. \\ \\
\indent At the outset, each BH is assigned a mass, spin, and age. We will illustrate the procedure with an example, and then summarize the procedure with a flowchart: Taking $\mu = 4\cross 10^{-11}$ eV, consider the evolution of a $5\ M_{\odot}$, $\chi = 0.95$ BH with an age of $10^8$ yrs. The superradiance condition, Eq$.$ \ref{eqn:SRineqdim}, determines which bound-state grows first.
\begin{equation}
    \begin{cases}
        \xi_{R} = 1.03 \ \ \ \xi_{\text{crit}} = 0.24 \ \ \ (m = 1) \\
        \xi_{R} = 0.76 \ \ \ \xi_{\text{crit}} = 0.48 \ \ \ (m = 2) \\
        \xi_{R} = 0.89 \ \ \ \xi_{\text{crit}} = 0.73 \ \ \ (m = 3) \\
        \xi_{R} = 0.94 \ \ \ \xi_{\text{crit}} = 0.97 \ \ \ (m = 4) \\
    \end{cases}
\end{equation}

\indent Since $\xi_{R} > \xi_{\text{crit}}$ for $m = 1, 2, 3$, the first superradiant bound-state is $n = 5,\ l = m = 4$, and it grows on a timescale of $\tau_{c} = 3.4$ yrs. The BH's mass and spin are decreased to $4.94\ M_{\odot}$ and $0.938$, respectively. Once the cloud has finished growing, it dissipates on a timescale $\tau_{\text{GW}} = 0.8$ yrs. The time from the BH's birth to the cloud's dissipation is only $\tau_{c} + \tau_{\text{GW}} = 4.2$ yrs, leaving plenty of time for new clouds to develop. We denote by $t_{r}$ the time remaining to the present. In this case, $t_{r} = 10^{8} - 4.2 \approx 10^{8}$ yrs.  \\ \\ 
\indent The next bound-state is $n = 6$ with $\tau_{c} = 3536$ yrs. The BH's mass and spin are decreased to $4.67\ M_{\odot}$ and $0.84$, respectively. Once the cloud has finished growing, it dissipates on a timescale $\tau_{\text{GW}} = 2445$ yrs. At this point, $t_{r} = 9.9994\cross10^{7}$ yrs -- still plenty of time left for further superradiance.  \\ \\ 
\indent The next (and final) bound-state is $n = 7$ with $\tau_{c} = 6\cross 10^{7}$ yrs. The BH's mass and spin are decreased to $4.5\ M_{\odot}$ and $0.74$, respectively. Once the cloud has finished growing, $t_{r} = 3.9\cross10^{7}$ yrs remain. The dissipation timescale $\tau_{\text{GW}} = 7\cross 10^{7}$ yrs. Since $\tau_{GW} > t_{r}$, the $n = 7$ cloud is still present today. It has an initial mass $M_{c} = 0.16\ M_{\odot}$, and it radiates at $f = 18.9$ kHz. Placing the source at $d = 1$ kpc (for example), the initial strain amplitude $h_{0} = 10^{-26}$ (Eq$.$ \ref{eqn:h0}). The signal observed today was emitted $d/c = 3300$ yrs ago, so the corresponding amplitude $h(t) = 6.9\cross10^{-27}$ (Eq$.$ \ref{eqn:h_of_t} with $t = t_{r} - d/c$). \\ \\
\indent Our simulation of the Galactic cloud population consists of applying the foregoing procedure to each of the BH's in the galaxy. If a given BH only permits a bound-state whose growth timescale is greater than the age of the universe ($\tau_{c} > \tau_{\text{uni}} = 1.38\times10^{10}$ yr), the host BH is removed from the simulation. \\ \\
\indent Our criterion for whether a given cloud is still present today is $\tau_{\text{GW}} > t_{r}$. For each black hole, there are only two final options: Either a cloud has finished growing and is still present today, or a cloud is growing on a timescale greater than the age of the universe. \\ \\
\indent Those BH's with an extant cloud are assigned a location in the Milky Way (Eqs$.$ \ref{rhodisks} and \ref{rhobulge}). Earth is assigned to an arbitrary, but fixed, point on the circle of radius $8.3$ kpc in the Galactic midplane. For a cloud located at distance $d$, we check the inequality $ct_{r} > d$ to determine if there has been enough time for GW's to propagate to Earth since the cloud formed. Those clouds for which $d > ct_{r}$ are presently unobservable, and we retain only those clouds for which $ct_{r} > d$. We summarize this section with the following flowchart: 
\begin{center}
    For a given $\mu$, $M_{\text{BH}}$, $\chi$, and BH age, find the lowest superradiant value of $n$. \newline
    
    $\rightarrow$ If $\tau_{c} > \tau_{\text{uni}}$, the BH is removed from the simulation. \newline

    $\rightarrow$ Otherwise, the dissipation timescale $\tau_{\text{GW}}$ determines whether a new cloud will start growing in accordance with $\tau_{\text{GW}} > t_{r}$ (cloud still present) or $\tau_{\text{GW}} < t_{r}$ (cloud has dissipated, and a new cloud begins growing).  \newline

    $\rightarrow$ Repeat the previous steps until one of two possibilities is obtained: a.) A cloud is growing with $\tau_{c} >$ the age of the universe, or b.) A cloud is still present \& radiating GW's today.  \newline

    $\rightarrow$ If the cloud hasn't dissipated yet, assign it a random position, and compute the GW strain at Earth's location only if the travel-time inequality $ct_{r} > d$ is true. \newline
\end{center}

%% file: Sections/Results.tex
\section{\label{sec:results} GW's from the axion cloud population}

\subsection{\label{subsec:populations} Cloud populations}

\begin{figure*}
    \includegraphics[width=0.7\linewidth]{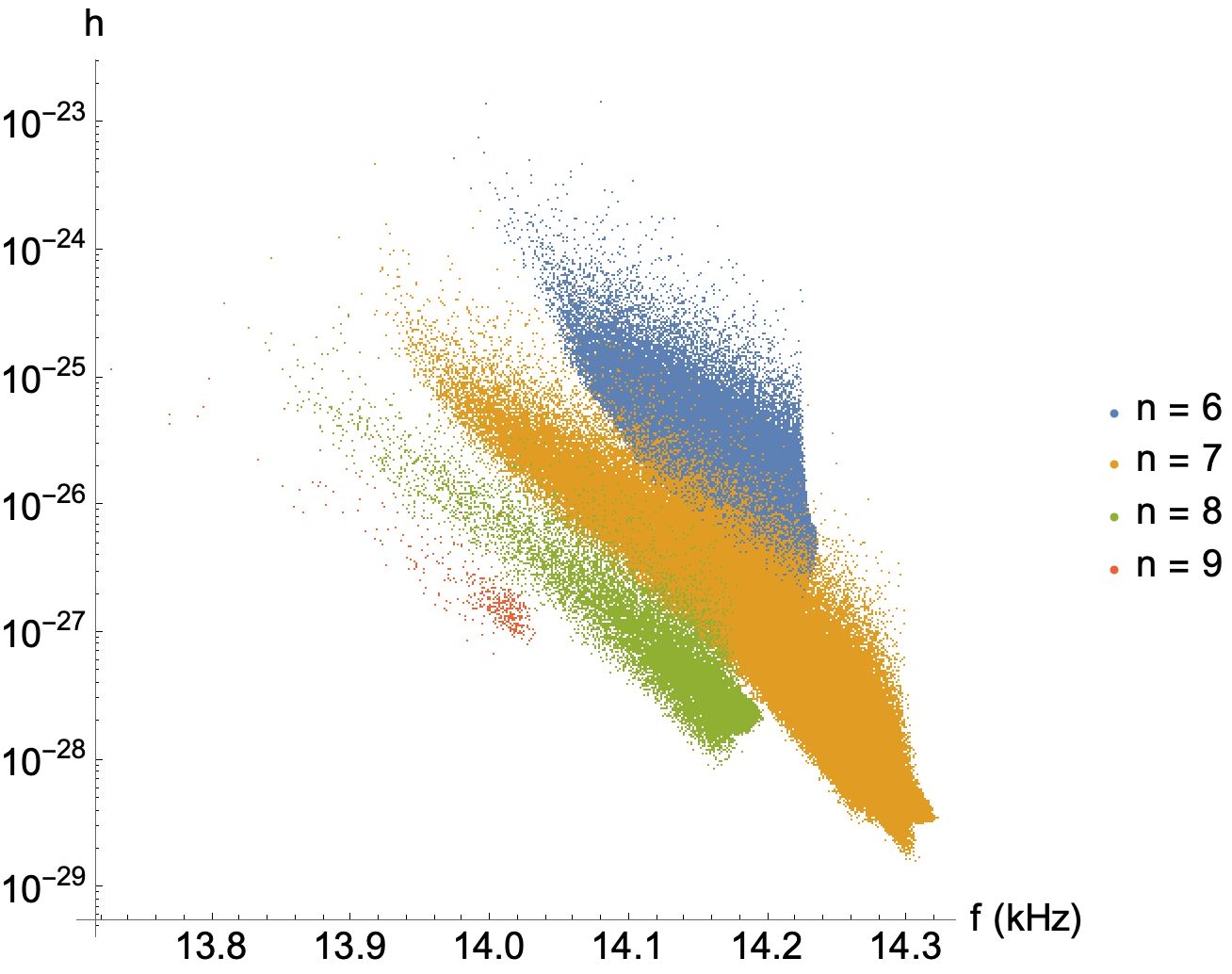}
    \caption{The population of axion clouds, with $\mu = 3\cross10^{-11}$ eV. From a population of $10^{8}$ BH's, the simulation returns $N_{c} = (9.32\pm0.03)\times10^{5}$ extant clouds. Of these, $8.3\times10^{5}$ ($89\%$) satisfy the GW travel-time condition (Sec$.$ V). Since GW amplitudes \& frequencies are, necessarily, only computed for those clouds satisfying the travel-time condition, it should be understood that only those clouds are represented in this figure, as well as in Figs$.$ \ref{fig:pop4e11} and \ref{fig:pop6e11}.} 
    \label{fig:pop3e11}
\end{figure*}

\begin{figure*}
    \includegraphics[width=0.7\linewidth]{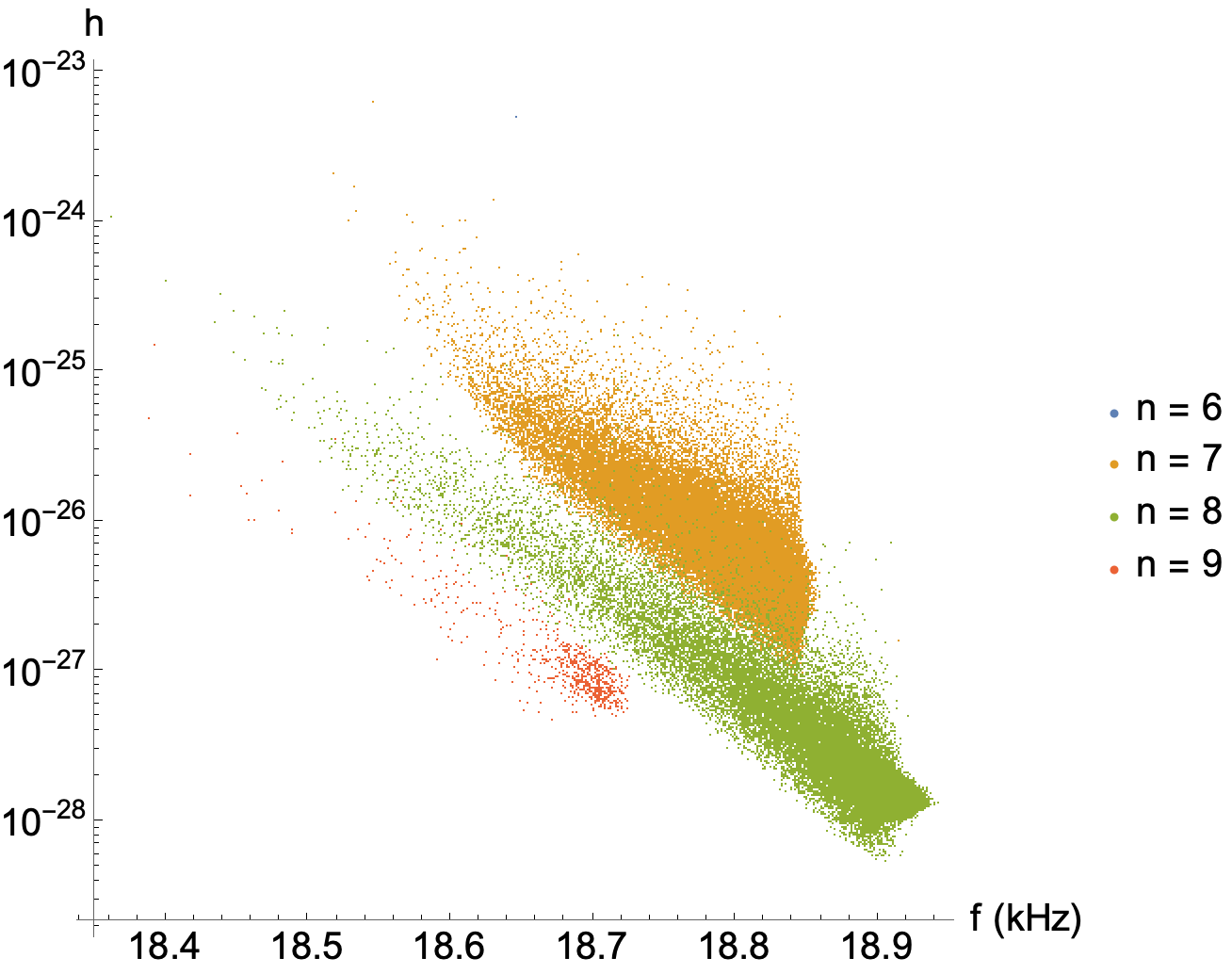}
    \caption{The population of axion clouds, with $\mu = 4\cross10^{-11}$ eV. From a population of $10^{8}$ BH's, the simulation returns $N_{c} = (1.17\pm0.01)\times10^{5}$ extant clouds. Of these, $9.5\times10^{4}$ ($81\%$) satisfy the GW travel-time condition.} 
    \label{fig:pop4e11}
\end{figure*}

\begin{figure*}
    \centering
    \includegraphics[width=0.7\textwidth]{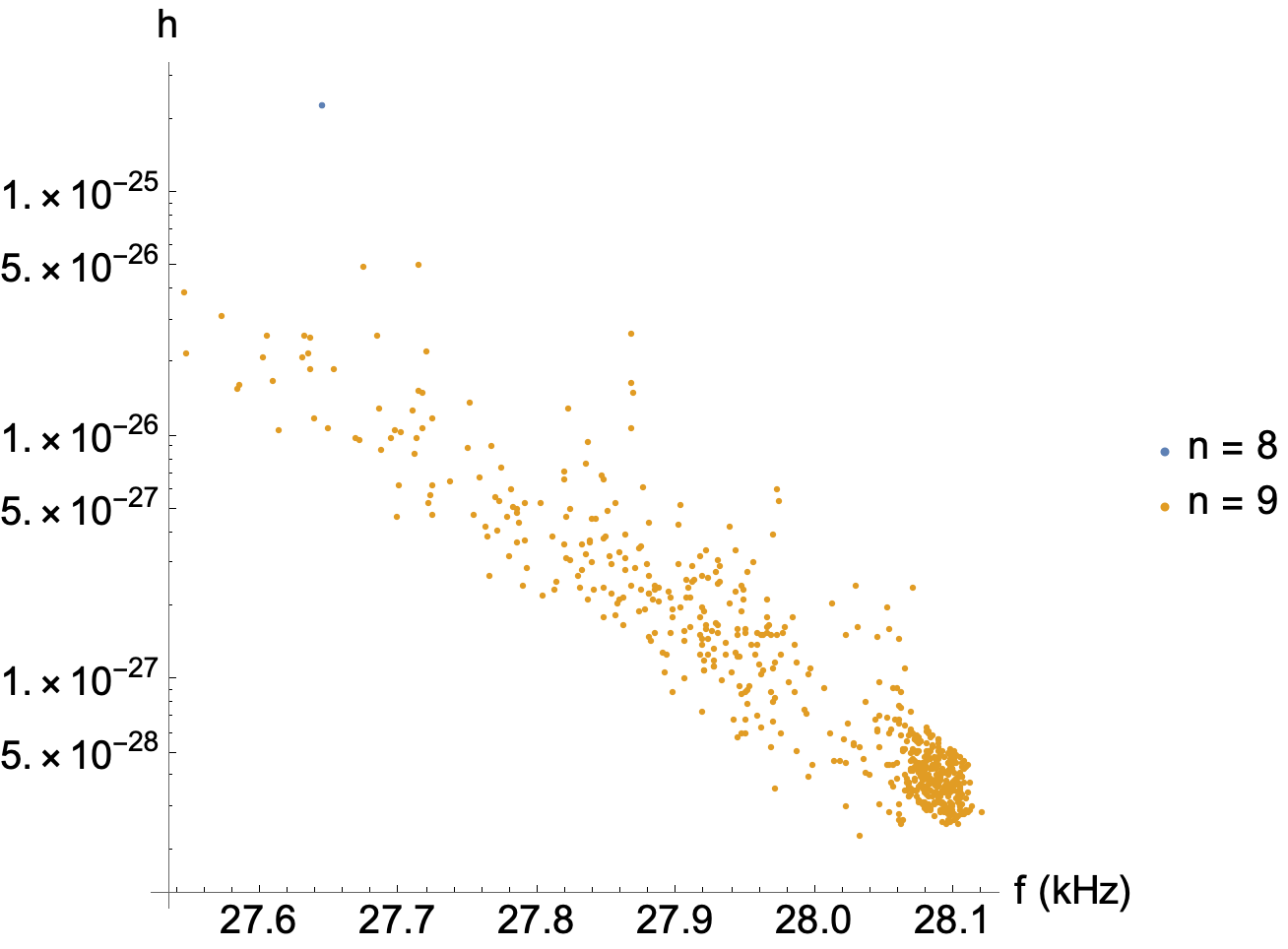}
    \caption{The population of axion clouds, with $\mu = 6\cross10^{-11}$ eV. From a population of $10^{8}$ BH's, the simulation returns $N_{c} = 900\pm200$ extant clouds. Of these, $620$ ($69\%$) satisfy the GW travel-time condition. The apparent gap in the $n=9$ band at $28$ kHz is an artifact due to the small number of clouds.} 
    \label{fig:pop6e11}
\end{figure*}

\begin{figure*}
    \centering
    \includegraphics[width=0.5\textwidth]{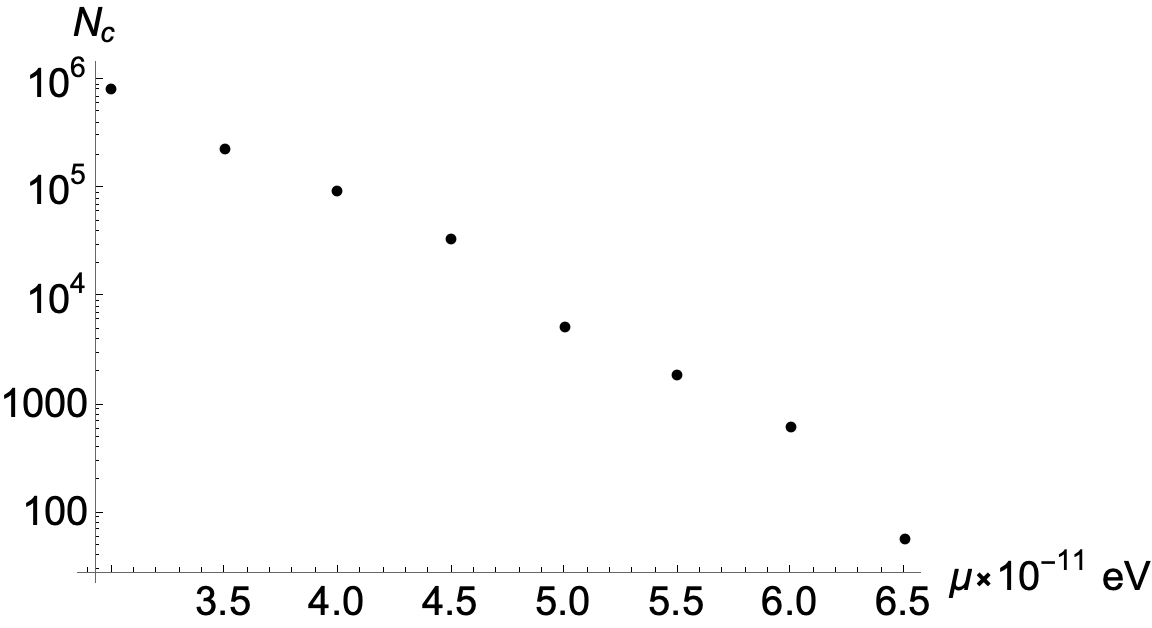}
    \caption{The number of extant axion clouds $N_{c}$ drops rapidly with increasing boson mass $\mu$, and it extrapolates to zero around $6.6\times 10^{-11}$ eV, corresponding to $f \approx 32$ kHz. In light of the lower-mass-gap objects found by LIGO-Virgo, we note that superradiant instabilities with a $\mathcal{O}(10^{-11})$ eV boson are stronger for $M_{\text{BH}} < 5\ M_{\odot}$ than for $M_{\text{BH}} > 5\ M_{\odot}$. Since the BH's involved in our simulations are of the latter type, the true number of clouds, for any boson mass, could be greater than our estimate by a factor depending on the mass distribution \& total number of lower-mass-gap BH's in the Milky Way.} 
    \label{fig:nclouds}
\end{figure*}

\begin{figure*}
    \centering
    \includegraphics[width=0.6\textwidth]{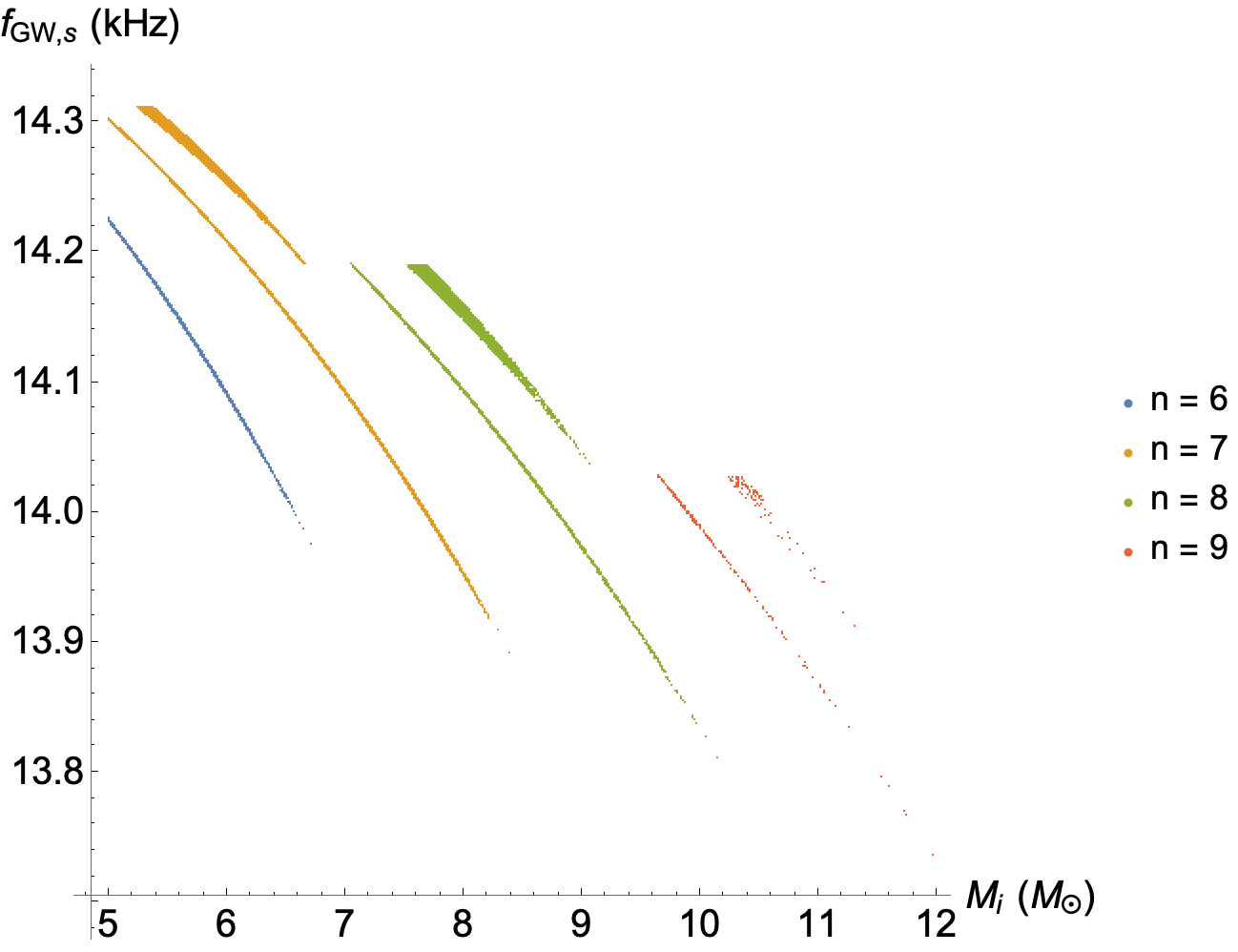}
    \caption{The source-frame GW frequencies $f_{\text{GW,s}}$ and initial BH masses $M_{i}$ for the extant clouds in the $3\times 10^{-11}$ eV simulation. The spread of frequencies below $f_{0} \propto 2\mu$ is a result of gravitational redshift.} 
    \label{fig:fmbh0}
\end{figure*}

\indent The total number of stellar-origin black holes has been estimated to be $\mathcal{O}(10^{8})$ from the Milky Way's supernova rate of $\mathcal{O}(1)\ \text{century}^{-1}$ \cite{2006Natur.439...45D}, and from population-synthesis estimates \cite{2020A&A...638A..94O}. We take $N_{\text{BH}} = 10^{8}$, bearing in mind that the true number could be larger by a factor of a few, or even another order-of-magnitude \cite{2002MNRAS.334..553A}. We have simulated the axion cloud population for $\mu = (3, 3.5, 4, 4.5, 5, 5.5, 6, 6.5)\times10^{-11}$ eV.  \\ \\
\indent The output of a simulation is a collection of all extant BH-cloud systems in the Milky Way. Those BH's which have experienced the growth of a single cloud are described by a list comprising the BH age, the initial and final values of the BH mass \& spin, the bound-state \{$n, l, m, \xi_{R}, \xi_{I}$\}, the cloud's properties -- mass $M_{c}$, growth timescale $\tau_{c}$, and dissipation timescale $\tau_{\text{GW}}$, -- the source distance $d$, and the GW frequency and amplitude $(f, h).$ BH's which have experienced the growth of multiple bound-states are each characterized by a set of such lists, one per bound-state. The GW frequency and amplitude are only computed for the extant cloud, all previous bound-states having already dissipated.  \\ \\
\indent For a given axion mass, the number of extant clouds is a random variable whose mean and standard deviation are estimated by performing twenty simulations with $5\times10^{6}$ BH's per simulation, computing the sample mean \& sample standard deviation of $N_{c}$ over the $20$ trials, and then multiplying them by $20$ and $\sqrt{20}$, respectively. \\ \\
\indent An ensemble of GW signals from axion clouds is a scatter plot in the $h\ \text{vs.}\ f$ plane, as in Figs$.$ \ref{fig:pop3e11}, \ref{fig:pop4e11}, and \ref{fig:pop6e11}$.$ The distribution of amplitudes and frequencies is not random, but consists of well-defined bands corresponding to the various occupied bound-states. The lowest bound-state resulting from our simulations is $n = 6$, reflecting the general difficulty for stellar-mass BH's to produce clouds in the LSD band.  \\ \\
\indent Also reflecting this difficulty is the rapid decline in the number of clouds $N_{c}$ with increasing boson mass $\mu$ (Fig$.$ \ref{fig:nclouds})$.$ For $\mu=3\cross10^{-11}$ eV, $N_{c} = (9.323\pm0.007)\times10^{5}$, while at $\mu=6.5\cross10^{-11}$ eV, the number has dropped to $130\pm10$. $N_{c}$ goes to zero around $6.6\times 10^{-11}$ eV, corresponding to a nominal upper limit of $\approx 32$ kHz for signals expected in the LSD band. Higher-frequency signals could occur from BH's with $M_{\text{BH}} < 5\ M_{\odot}$, especially in light of the recent discoveries of lower-mass-gap objects. \\ \\
\indent In all cases, the distribution of GW frequencies occurs below the nominal value $f_{0} \propto 2\mu$ due to the positive scaling of gravitational redshift with BH mass. This interpretation is confirmed by a plot of the source-frame GW frequencies $f_{\text{GW,s}}$ and initial BH masses $M_{i}$ for all extant clouds in a given simulation (Fig$.$ \ref{fig:fmbh0}, with $\mu = 3\cross10^{-11}$ eV). For each scalar bound-state, there is a tight relationship, with more massive BH's producing lower-frequency clouds. \\ \\
\indent In the introduction (Sec$.$ \ref{sec:level1}), we noted a potential connection between the QCD axion and the GUT scale $\Lambda_{\text{GUT}}$ (Eq$.$ \ref{eqn:PQmass}): An axion of mass $\mathcal{O}(10^{-10})$ eV corresponds to $f_{a} \approx \Lambda_{\text{GUT}}$. If the solution to the strong-CP problem is tied to GUT phenomenology, then discovery of an $\mathcal{O}(10^{-10})$ eV axion would be an exciting, albeit indirect, form of evidence for grand unification. The number of clouds in the Milky Way dropping to zero around $6.6\times 10^{-11}$ eV would seem to preclude the possibility of detecting an $\mathcal{O}(10^{-10})$ eV axion -- and, by extent, of probing GUT-scale physics with the LSD. Lower-mass-gap BH's could produce clouds at higher $\mu$, thereby reviving hopes of finding a GUT-scale axion. Another possibility is that $\Lambda_{\text{GUT}}$ is model-dependent, giving rise to a range of possible values including $10^{17}$ GeV, which corresponds to $\mathcal{O}(10^{-11})$ eV bosons.

\subsection{\label{subsec:resolvable} Resolvable signals}

\begin{figure*}
    \centering
    \includegraphics[width=0.7\textwidth]{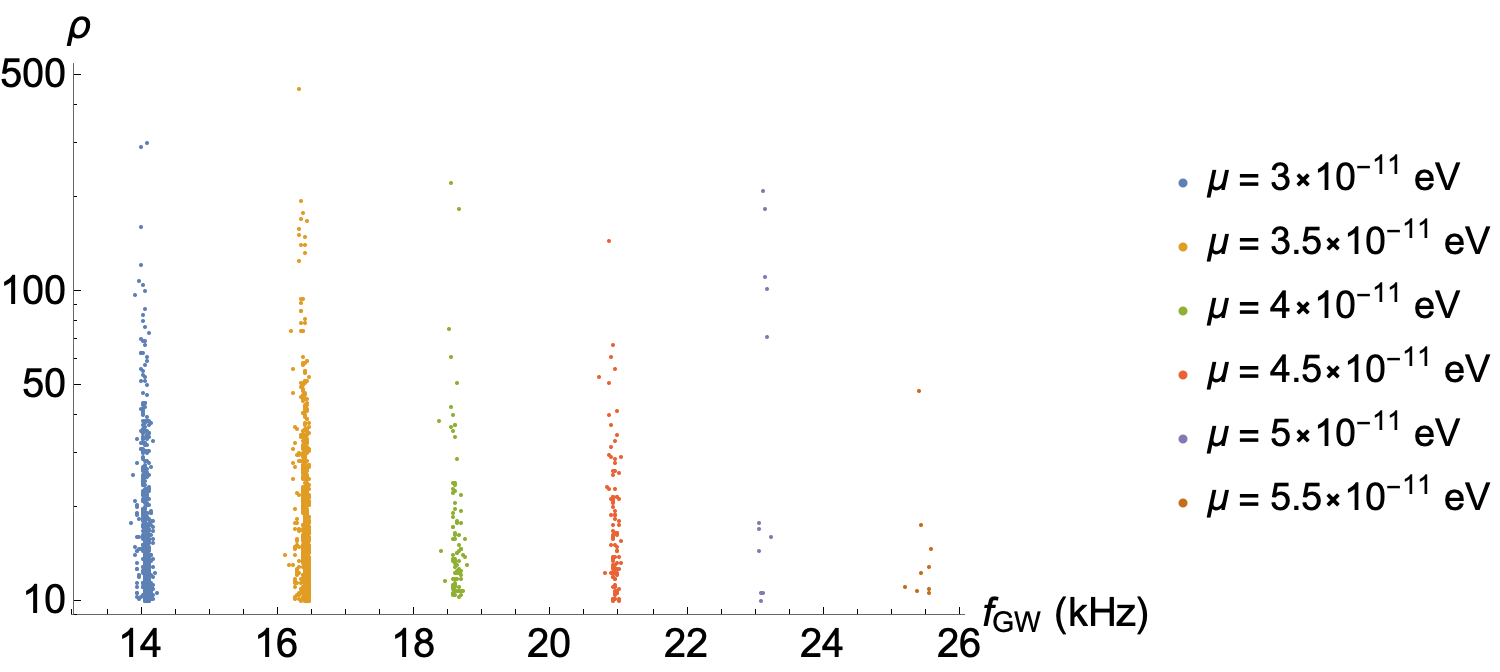}
    \caption{The SNR's $\rho$ of resolvable GW signals from simulated axion cloud populations, for the $100$-m LSD. As a function of $\mu$, the number of resolvable signals drops to zero beyond $5.5\times 10^{-11}$ eV, thereby constraining the expected frequency range of resolvable sources to $f < 26$ kHz.} 
    \label{fig:resolvable}
\end{figure*}

\begin{figure*}
    \centering
    \includegraphics[width=0.7\textwidth]{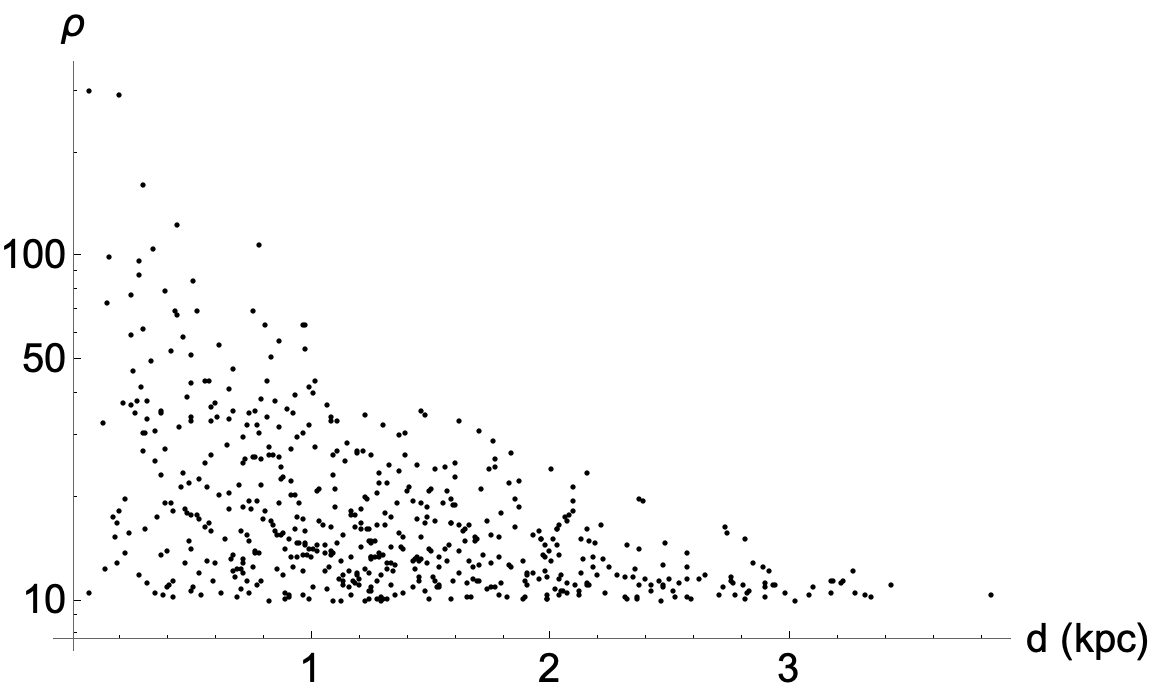}
    \caption{Scatter plot of the SNR's $\rho$ and source distances $d$ of resolvable signals for the $100$-m LSD, taking $\mu=3\times 10^{-11}$ eV. There appears to be an upper limit $d\lesssim3$ kpc, and certainly a hard upper limit $d<4$ kpc, suggesting a way to distinguish between spin-$0$ and spin-$1$ bosonic clouds: Proca clouds are generally louder GW sources than axion clouds, so a GW signal with $d\gg3$ kpc could be taken as evidence of a spin-$1$ field.} 
    \label{fig:resolvabledist}
\end{figure*}

\indent The standard result for coherent detection of a continuous monochromatic signal, $h(t) = h_{0}\cos (\omega t)$, is that the signal-to-noise ratio (SNR) $\rho$ grows as the square-root of the coherent integration time $T_{\text{coh}}$ \cite{2021LRR....24....4A}
\begin{equation}
    \label{eqn:snr}
    \rho = \frac{h_{0}\sqrt{T_{\text{coh}}}}{\sqrt{S_{n}(f)}}
\end{equation}
where $\sqrt{S_{n}(f)}$ is the one-sided amplitude spectral density (ASD) of the detector noise (the `sensitivity curve') evaluated at the GW frequency, and the trapping frequency of the levitated sensor is constant during the entire observation time. The LSD is an Earth-bound detector for which the observed signal, Eq$.$ \ref{ht}, experiences both amplitude modulation via the angular-dependence of the sensitivity, and phase modulation via the Earth's daily (diurnal) rotation and orbital motion. These corrections require knowledge of both the Earth's position and the source position to high accuracy. In an all-sky blind search for axion clouds, the source position is not known ahead of time, so a realistic search will require a large number of templates corresponding to many sky locations. A coherent search for $\mathcal{O}(10)$ kHz GW's over the full observation time (`fully-coherent search') of four months is not computationally-feasible because of the need to take fine steps in parameter space. A `semi-coherent' search, in which the observation time is divided into several segments, sacrifices sensitivity for a great reduction in the number of templates needed to perform a coherent search on each segment \cite{2023LRR....26....3R}. \\ \\
\indent As a means of setting upper limits on the expected number of resolvable signals, we compute the SNR for the idealized case of a detector freely orbiting the Milky Way at the same radius as the Solar System (i.e$.$ not attached to a planet or star system). There is a positive frequency derivative due to the gradual dissipation of the cloud which, however, is estimated to be too small to be detected \cite{2020PhRvD.102f3020Z}. The signal remains perfectly coherent over the full observation time, and a hypothetical search performed at the Solar System barycenter would simply involve Fourier-transforming the data and looking for lines in the power spectrum. \\ \\ 
\indent Taking $T_{\text{coh}} = 10^{7}$ s, and with the projected sensitivity curves for the current $1$-m LSD prototype, as well as for future $10$-m and $100$-m versions \cite{2022PhRvL.128k1101A}, we compute the corresponding SNR's for all sources in the galaxy. We count those with $\rho > \rho_{\text{t}}$ as resolvable, and we adopt the threshold $\rho_{\text{t}} = 10$ (Fig$.$ \ref{fig:resolvable}). \\ \\
\indent The `loudness' of a signal is determined primarily by the source distance. The distance, in turn, is a random variable determined by the randomly-assigned position vector (Eqs$.$ \ref{rhodisks} and \ref{rhobulge}) of the source. Thus, for a given set of extant clouds, the number of individually-resolved sources $N_{\text{res}}$ will vary each time we re-assign their position vectors. We estimate the mean \& standard deviation of $N_{\text{res}}$ for a given population of extant clouds by laying them down in the Galaxy $N_{\text{reshuffle}} = 100$ times and counting how many are resolvable in each `re-shuffling'. The mean \& standard deviation are then computed as
\begin{equation}
    \Bar{N}_{\text{res}} = \frac{1}{N_{\text{reshuffle}}}\sum_{i = 1}^{N_{\text{reshuffle}}}N_{\text{res}, i}
\end{equation}
\begin{equation}
    \Bar{\sigma}_{\text{res}} = \frac{1}{N_{\text{reshuffle}} - 1}\sum_{i = 1}^{N_{\text{reshuffle}}}\big(N_{\text{res}, i} - \Bar{N}_{\text{res}}\big)^{2}
\end{equation}
\indent With a $100$-m detector, assuming $\mu = 3\times 10^{-11}$ eV, $\Bar{N}_{\text{res}} = 600$ with $\Bar{\sigma}_{\text{res}} = 20$. In the most pessimistic case ($\mu = 5.5\times 10^{-11}$ eV), there are only $\mathcal{O}(1)$ resolvable signals, and we have not estimated the associated uncertainty. The $10 - 26$ kHz range is where we expect resolvable signals to be present for a $100$-m LSD. For a $10$-m instrument, $\mathcal{O}(1)$ resolvable signals appear at $\mu = 3\times 10^{-11}$ eV, while a $1$-m instrument does not have the required sensitivity to detect individual sources. \\ \\
\indent In the event a continuous monochromatic signal is detected by the LSD, we will have to answer the question: Is this signal from an `axion cloud' -- a superradiant bound-state of a scalar (spin-0) field -- or from a cloud involving a spin-1 (`Proca') field? In general, Proca fields give rise to stronger GW signals than scalar fields \cite{2020PhRvD.101b4019S}. As a result, we would expect resolvable signals from Proca clouds to be found at greater distances than those from scalar clouds. For the $100$-m detector, with $\mu = 3\times10^{-11}$ eV, the resolvable signals are depicted in terms of their SNR's and source distances in Fig$.$ \ref{fig:resolvabledist}. The vast majority are less than $3$ kpc away. Turning this on its head, the detection of a continuous monochromatic signal with an inferred distance significantly greater than $3$ kpc could be a potential indicator of a spin-1 field.
\begin{center}
    \subsection{\label{subsec:foreground} Unresolved signals}
\end{center}

\begin{figure*}
    \centering
    \includegraphics[width=0.6\textwidth]{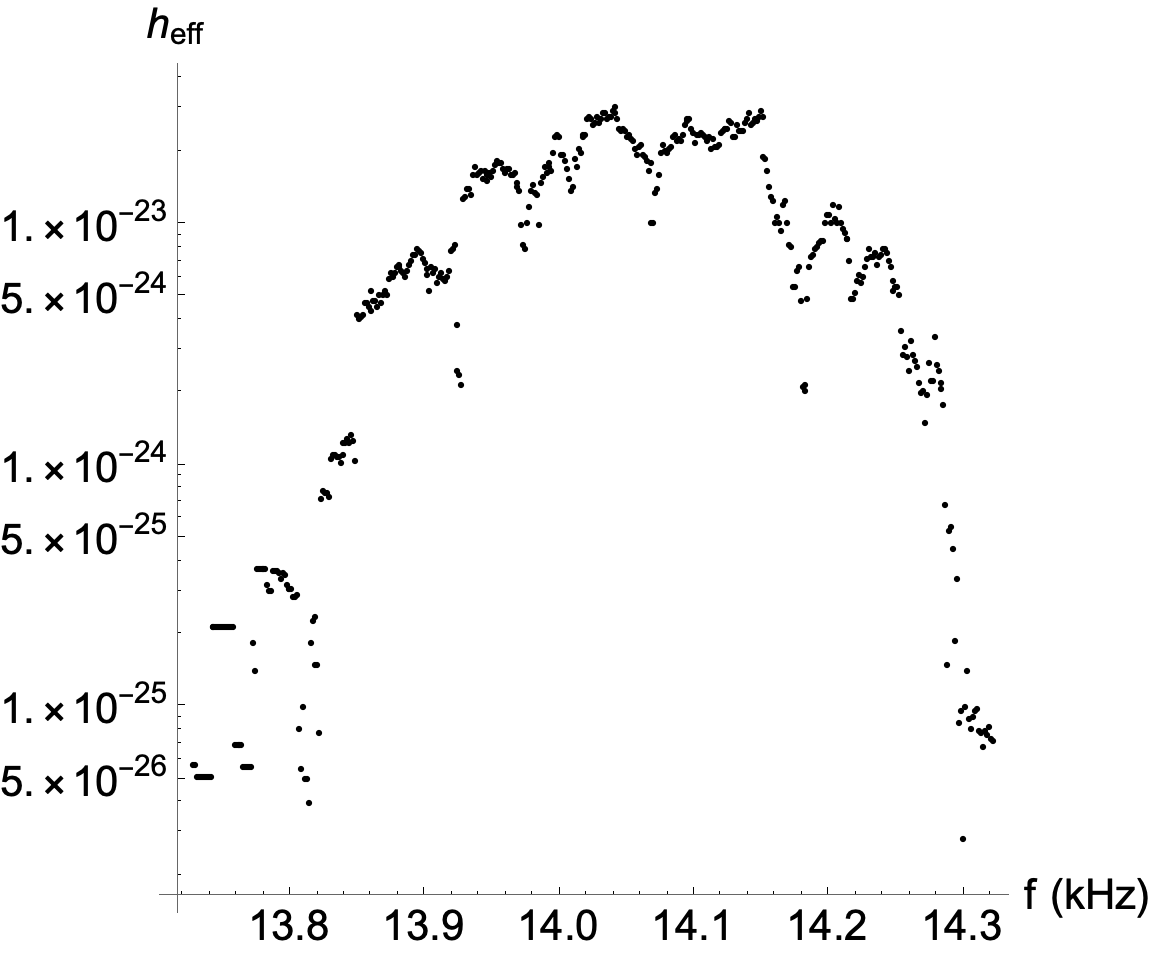}
    \caption{The effective amplitudes $h_{\text{eff}}$, as defined in Eq$.$ \ref{eqn:hq}, for the binned GW amplitudes of a simulated population of axion clouds with $\mu = 3\times10^{-11}$ eV. The width of each bin is a factor $10^{-2}$ of the central frequency $f_{c}$, reflecting the FWHM of the detector response when the trap frequency is $f_{c}$.} 
    \label{fig:heff3}
\end{figure*}

\begin{figure*}
    \centering
    \includegraphics[width=0.7\textwidth]{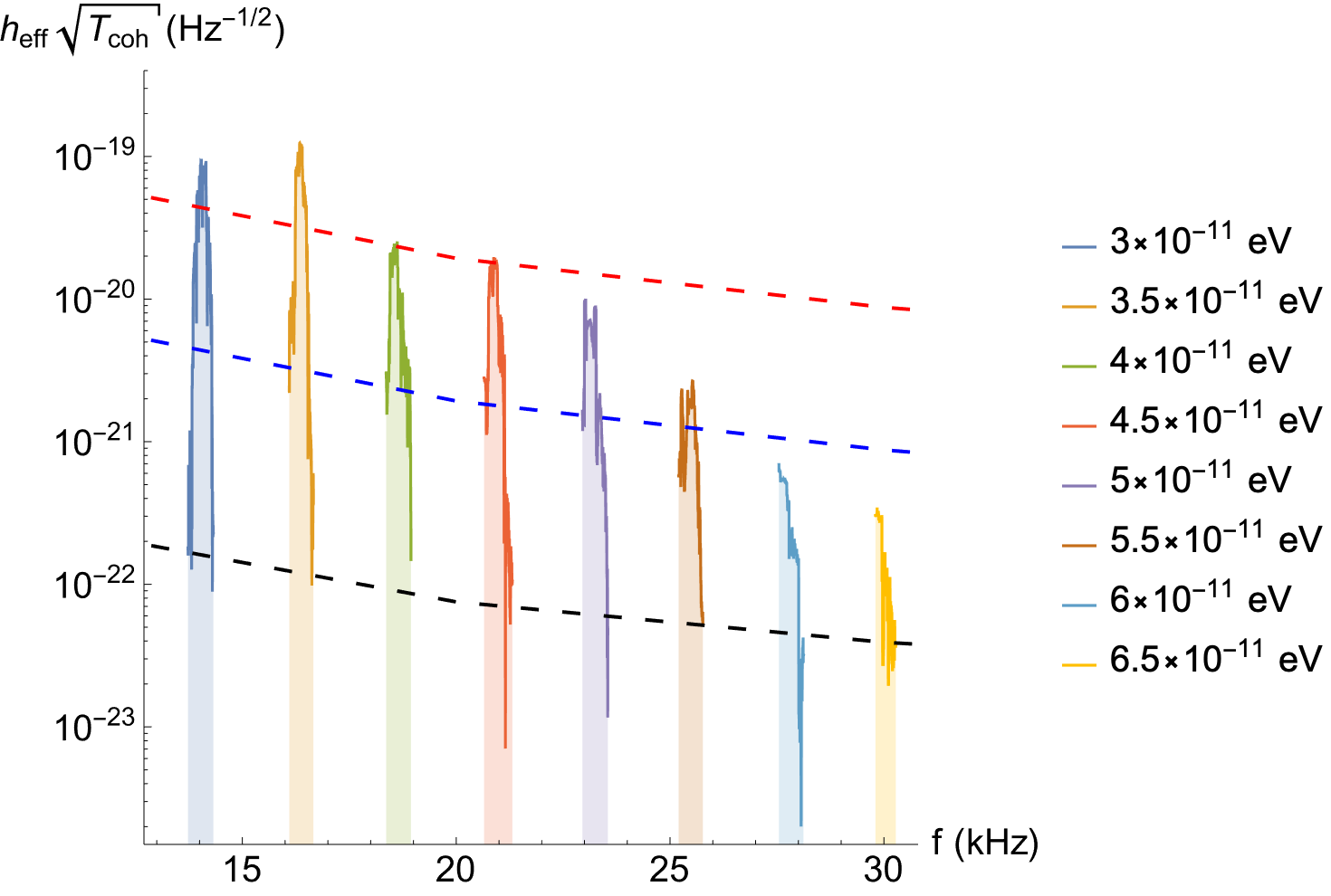}
    \caption{The confusion-limited foreground of unresolved axion clouds, as a function of the boson mass $\mu$, taking $T_{\text{coh}} = 10^{7}$ s. The $1-$, $10-$, and $100-$m LSD sensitivity curves are depicted by the red, blue, and black dashed curves, respectively.} 
    \label{fig:confusion}
\end{figure*}

\begin{figure*}
    \centering
    \includegraphics[width=0.6\textwidth]{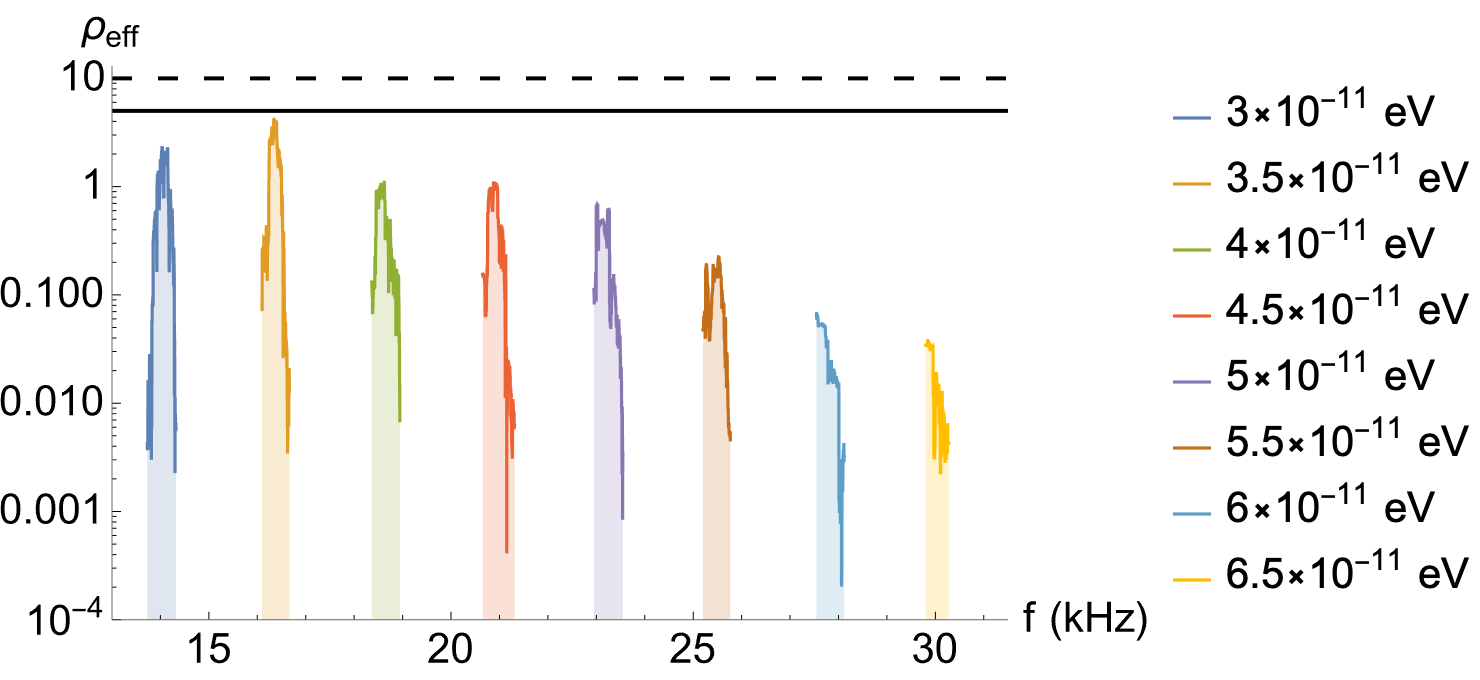}
    \caption{The corresponding SNR $\rho_{\text{eff}}$ of the confusion-limited foreground with respect to the $1$-m LSD sensitivity curve. The dashed line depicts the detection threshold $\rho_{t} = 10$ adopted in this paper, and the solid line represents a lower threshold of $5$. The foreground does not rise above either threshold.} 
    \label{fig:cfrho1}
\end{figure*}

\begin{figure*}
    \centering
    \includegraphics[width=0.6\textwidth]{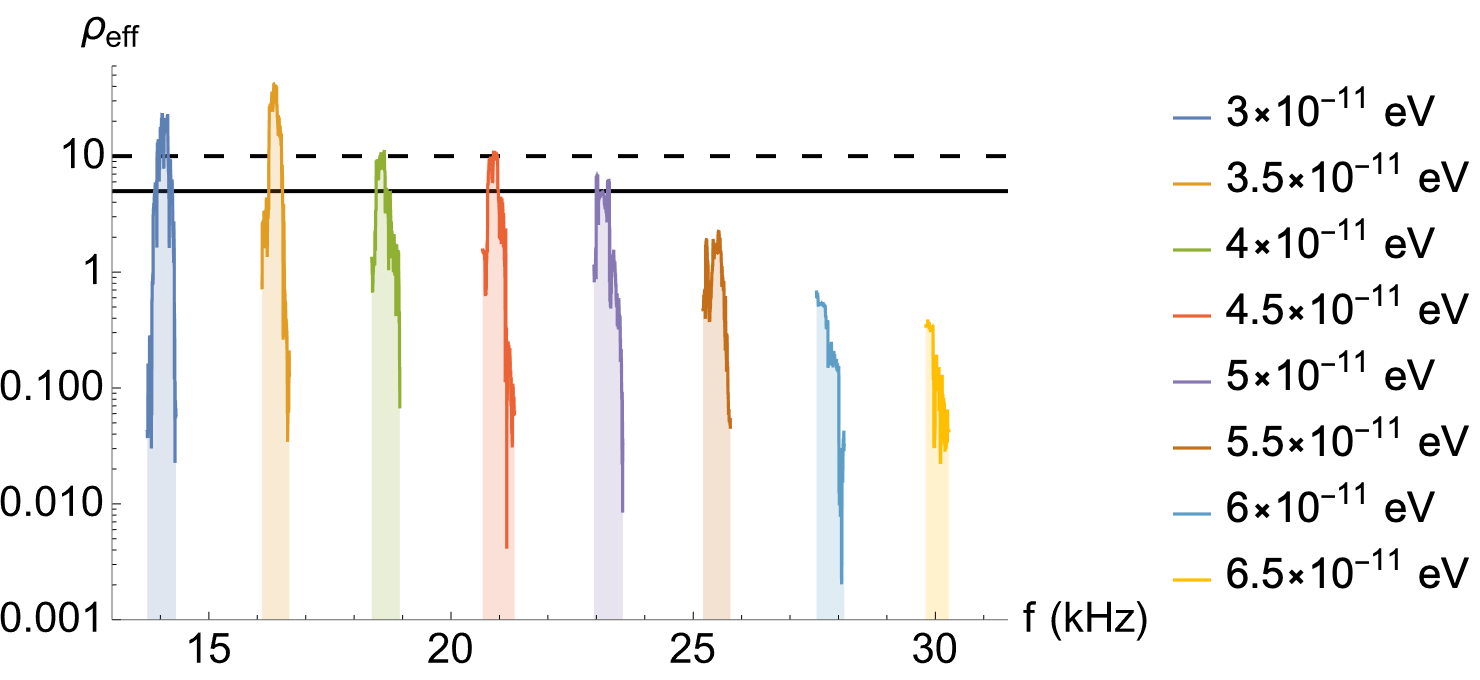}
    \caption{The corresponding SNR $\rho_{\text{eff}}$ of the confusion-limited foreground with respect to the $10$-m LSD sensitivity curve. The dashed line depicts the detection threshold $\rho_{t} = 10$ adopted in this paper, and the solid line represents a lower threshold of $5$. Over an observation time of $10^{7}$ s, the $10$-m instrument has the required sensitivity to detect the foreground in the range $(3-4.5)\times10^{-11}$ eV, although in the $(4-4.5)\times10^{-11}$ eV range, only the peak rises above the threshold, and only barely so. If a lower threshold $\rho_{t} = 5$ were adopted, the range of boson masses could be extended up to $5\times10^{-11}$ eV.} 
    \label{fig:cfrho10}
\end{figure*}

\begin{figure*}
    \centering
    \includegraphics[width=0.6\textwidth]{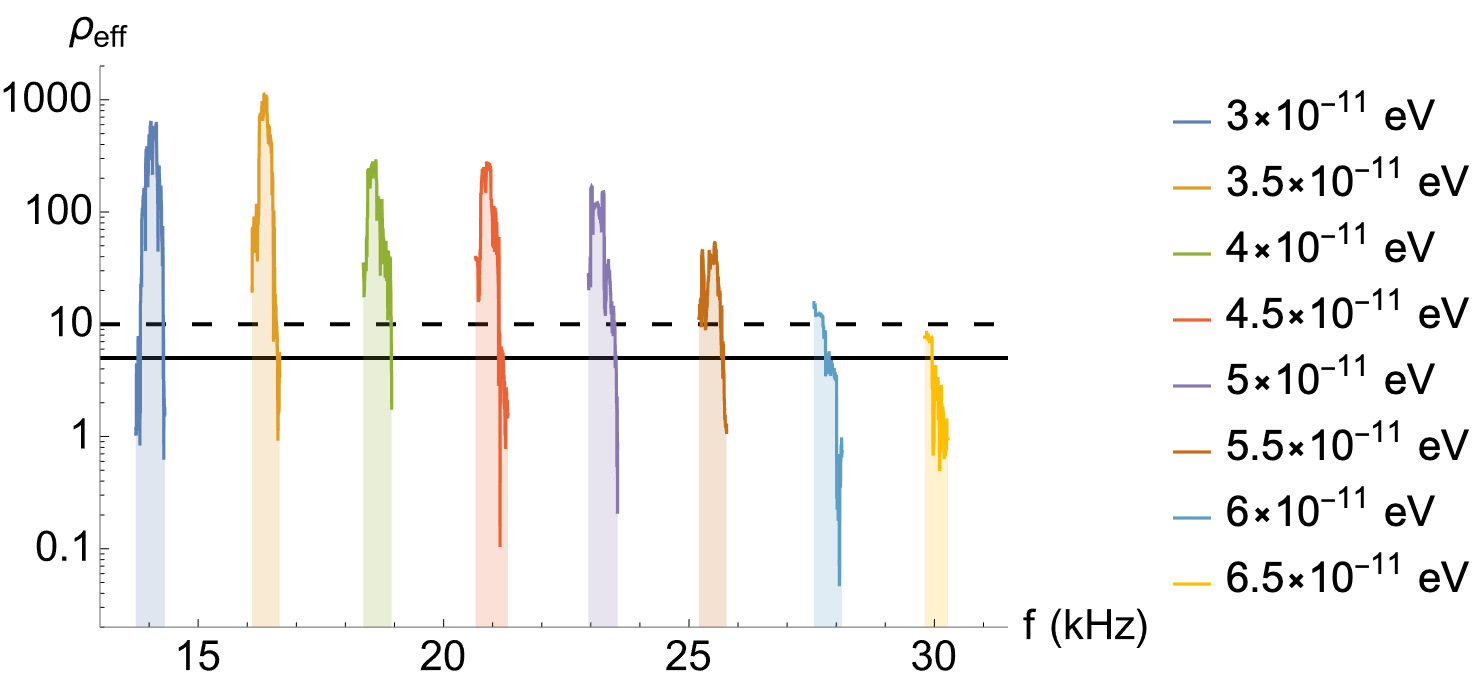}
    \caption{The corresponding SNR $\rho_{\text{eff}}$ of the confusion-limited foreground with respect to the $100$-m LSD sensitivity curve. The dashed line depicts the detection threshold $\rho_{t} = 10$ adopted in this paper, and the solid line represents a lower threshold of $5$. In the $(3-6)\times10^{-11}$ eV range, the foreground is predicted to be detectable by a $100$-m instrument with effective SNR's upwards of a thousand in the $(3-3.5)\times10^{-11}$ eV range.} 
    \label{fig:cfrho100}
\end{figure*}

\indent For all boson masses, the majority of GW signals have amplitudes less than $10^{-23}$, with the weakest having $h = \mathcal{O}(10^{-29})$. The unresolvable signals incoherently combine to form a Galactic confusion foreground which manifests as an excess noise in the detector. As before, we neglect the diurnal and annual modulations of the background, and instead provide a preliminary estimate of the foreground's strength compared to the nominal $1$-m, $10$-m, and $100$-m  LSD sensitivity curves. In a strain-frequency plot (e.g$.$ Fig$.$ \ref{fig:pop3e11}), we bin the cloud amplitudes (with bin width $\delta f = 10^{-2}f_{c}$, where $f_{c}$ is the center frequency of a given bin, and the factor $10^{-2}$ is the full-width-at-half-maximum (FWHM) of the trapped object's response function around $f_{c}$), and we associate an rms amplitude, defined as follows, with each bin. \\ \\
\indent We start by creating a bin centered on the frequency of the cloud with the smallest GW frequency in a strain-frequency plot, e.g. Fig$.$ \ref{fig:pop3e11}$.$ All axion clouds emit monochromatic signals,
\begin{equation}
    \label{eqn:hi}
    h_{i}(t) = h_{0,i}\cos\big(2\pi f t + \phi_{i}\big)
\end{equation}
where the phases $\phi_{i}$ are uniformly-distributed between $0$ and $2\pi$, and $i$ runs over all clouds in the bin. The squared sum of all signals in the bin is time-averaged over a period $T_{c} = 1/f_{c}$, where $f_{c}$ is the frequency at the center of the bin; The result is a dimensionless time-averaged power associated with that bin. The square-root of the power represents an effective amplitude $h_{\text{eff}}$ of the confusion foreground in the bin,
\begin{equation}
    \label{eqn:hq}
    h_{\text{eff}} \vcentcolon= \sqrt{\frac{1}{T_{c}}\int_{0}^{T_{c}}dt \bigg[\sum_{i}h_{i}(t) \bigg]^{2}}
\end{equation}
\indent We then create a new bin with center frequency $f_{c} \big|_{\text{new}}$ and width $\delta f\big|_{\text{new}}$,
\begin{equation}
    f_{c} \big|_{\text{new}} = f_{c}\big|_{\text{old}} + 10^{-2}\delta f\big|_{\text{old}}
\end{equation}
\begin{equation}
    \delta f\big|_{\text{new}} = 10^{-2}f_{c}\big|_{\text{new}}
\end{equation}
and we compute $h_{\text{eff}}$ for this bin. The center frequency is shifted rightwards by a fraction (arbitrarily chosen to be $10^{-2}$) of the previous bin width so that adjacent bins overlap, ensuring some degree of continuity in $h_{\text{eff}}$ vs. $f$. We continue until we reach the rightmost end of the cloud population. Each bin is then characterized by an ordered pair $(f_{c},\ h_{\text{eff}})$ (Fig$.$ \ref{fig:heff3}).  \\ \\
\indent A preliminary method for estimating the LSD's sensitivity to the confusion foreground is to treat each pair $(f_{c},\ h_{\text{eff}})$ as if they were the frequency and amplitude of a hypothetical monochromatic signal whose corresponding effective SNR $\rho_{\text{eff}}$, computed via Eq$.$ \ref{eqn:snr}, is then compared to a threshold $\rho_{\text{t}}$. We continue to require $\rho_{\text{t}} = 10$. The numerator and denominator of Eq$.$ \ref{eqn:snr} $\big(h_{\text{eff}}\sqrt{T_{\text{coh}}}\ \text{and}\ \sqrt{S_{n}(f_{c})},\ \text{respectively}\big)$ are shown separately in Fig$.$ \ref{fig:confusion}, and their ratio (the SNR) is shown in Figs$.$ \ref{fig:cfrho1}, \ref{fig:cfrho10}, and \ref{fig:cfrho100} for the $1-$, $10-$, and $100-$m instruments, respectively. \\ \\
\indent We find that a single $1$-m LSD does not appear to have the required sensitivity to detect the foreground for any value of $\mu$. A $10$-m detector could detect the foreground with $\rho_{\text{eff}} = \mathcal{O}(10)$ if the axion mass $\mu \in (3-4)\times10^{-11}$ eV, while in the $(4-4.5)\times10^{-11}$ eV range, only the peak of the foreground rises to the threshold, and just barely so (Fig$.$ \ref{fig:cfrho10}). A 100-m instrument could detect the foreground with large $\rho_{\text{eff}}$ if $\mu \in (3-6)\times10^{-11}$ eV (Fig$.$ \ref{fig:cfrho100}). In the range $(3-3.5)\times10^{-11}$ eV, the peak value of $\rho_{\text{eff}}$ is $\mathcal{O}(10^{3})$, and remains $\mathcal{O}(10^{2})$ up to $5\times 10^{-11}$ eV. \\ \\

%% file: Sections/Conclusion.tex
\section{\label{sec:conclusion}Conclusion}
We have produced Galactic-scale populations of the hypothetical GW sources known as `axion clouds' with the axion mass chosen to correspond to frequencies in the $10-100$ kHz band. By computing superradiant bound-states up to $n = 9$, we have accounted for nearly all clouds with growth timescales less than the age of the universe. \\ \\
\indent The largest number of clouds occurs for the lightest boson mass capable of producing GW's at the frequencies of interest. This was to be expected, as superradiance occurs more readily for small $\alpha \propto \mu M_{\text{BH}}$. For a BH of mass $M \geq 5\ M_{\odot}$, the smallest value of $\alpha$ is obtained with the smallest allowed boson mass, $3\times10^{-11}$ eV. In this most optimistic case, the total number of extant clouds is close to $1$ million.   \\ \\ 
\indent The population of axion clouds has been assumed to be spatially-distributed within the Milky Way in the same way as the stellar disks and central bulge. Statistically, some may be near enough that the continuous monochromatic signal can be detected by observing over a long enough period of time, e.g$.$ $10^{7}$ s, such that the SNR rises above a given threshold $\rho_{\text{t}}$; We have imposed a stringent threshold $\rho_{\text{t}} = 10$, but we leave it for future work to determine the most appropriate threshold for our search pipeline. For a $100$-m instrument, several hundred resolvable signals are predicted to occur if $\mu \approx 3\times10^{-11}$ eV, but this number could be upwards of an order-of-magnitude larger if the total number of stellar-origin BH's is also larger than we have assumed (see the comment made at the beginning of Sec$.$ VI). For a $10$-m detector, only $\mathcal{O}(1)$ resolvable signals occur in our simulation at $\mu = 3\times10^{-11}$ eV. \\ \\
\indent Meanwhile, the ensemble of unresolved signals produces a confusion foreground which is estimated to be detectable with potentially large SNR by a $100$-m LSD, assuming $\mu \in \big(3-6\big)\times10^{-11}$ eV, or by a $10$-m instrument at moderate SNR, assuming $\mu \in \big(3-4.5\big)\times10^{-11}$ eV. \\ \\
\indent Finally, we note the following limitations of this work, as well as directions for future work: First, since isolated BH's have no EM counterpart, we do not know, ahead of time, the direction to these GW sources. Targeted \& directed searches for axion clouds will, therefore, not be possible for isolated BH's, and we must resort to blind all-sky searches. Doppler modulations of the GW frequency can be accounted for by including the source's right ascension \& declination as additional parameters in the signal model for Bayesian parameter estimation. To avoid the number of demodulation templates becoming prohibitively large, -- we do not want the time required for data analysis to be greater than the four-month observation period -- we will resort to semi-coherent techniques for constructing a detection statistic. A final complication might be that a source has a transverse proper motion large enough to change its sky location within the observation time. \\ \\
\indent In this work, we have used the SNR as a baseline detection statistic. Since we have not yet developed the full semi-coherent search pipeline, we do not yet know what is the most appropriate detection statistic for continuous monochromatic signals. Another outstanding question pertains to our scan strategy: Given that we will take data at each trapping frequency for 4 months, how finely should the set of frequencies be discretized? At what frequency should we begin our searches? Figs$.$ \ref{fig:resolvable} and \ref{fig:confusion} both indicate the $14 - 20$ kHz range to be the most optimistic, in terms of the expected number of resolvable signals and the strength of the confusion foreground, respectively.  \\ \\
\indent Additionally, our treatment of the confusion foreground has not accounted for the intrinsic anisotropy of the signal: The axion clouds will be distributed throughout the disks and bulge of the Milky Way, so the strength of the foreground will vary over the sky in a complicated way. Searches for stochastic signals typically involve an `excess-power' method, as well as cross-correlation between multiple detectors. Plans to build a second $1$-m instrument at UC Davis (in addition to the Northwestern detector) are in development, so while a single $1$-m detector might not have the requisite sensitivity, the prospects for a two- or multi-detector scheme are an exciting avenue of future study.

%% file: Sections/Appendix.tex
\appendix*

\section{Superradiant bound-states}
\begin{figure*}
    \centering
    \includegraphics[width=0.8\textwidth]{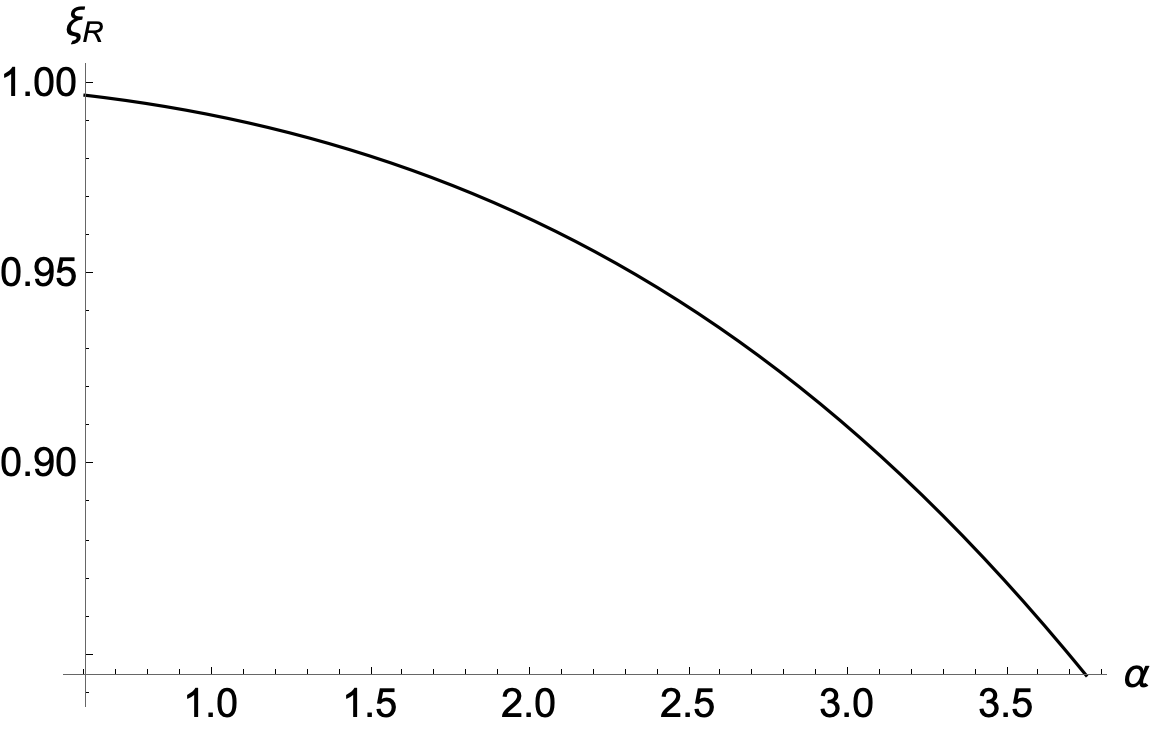}
    \caption{The real part $\xi_{R}$ of the $n = 8$ bound-state, for BH spin $\chi = 0.995$, plotted up to the associated maximum superradiant value of the coupling parameter, $\alpha_{\text{max}} = 3.75$.} 
    \label{fig:xir81}
\end{figure*}

\begin{figure*}
    \centering
    \includegraphics[width=1.0\textwidth]{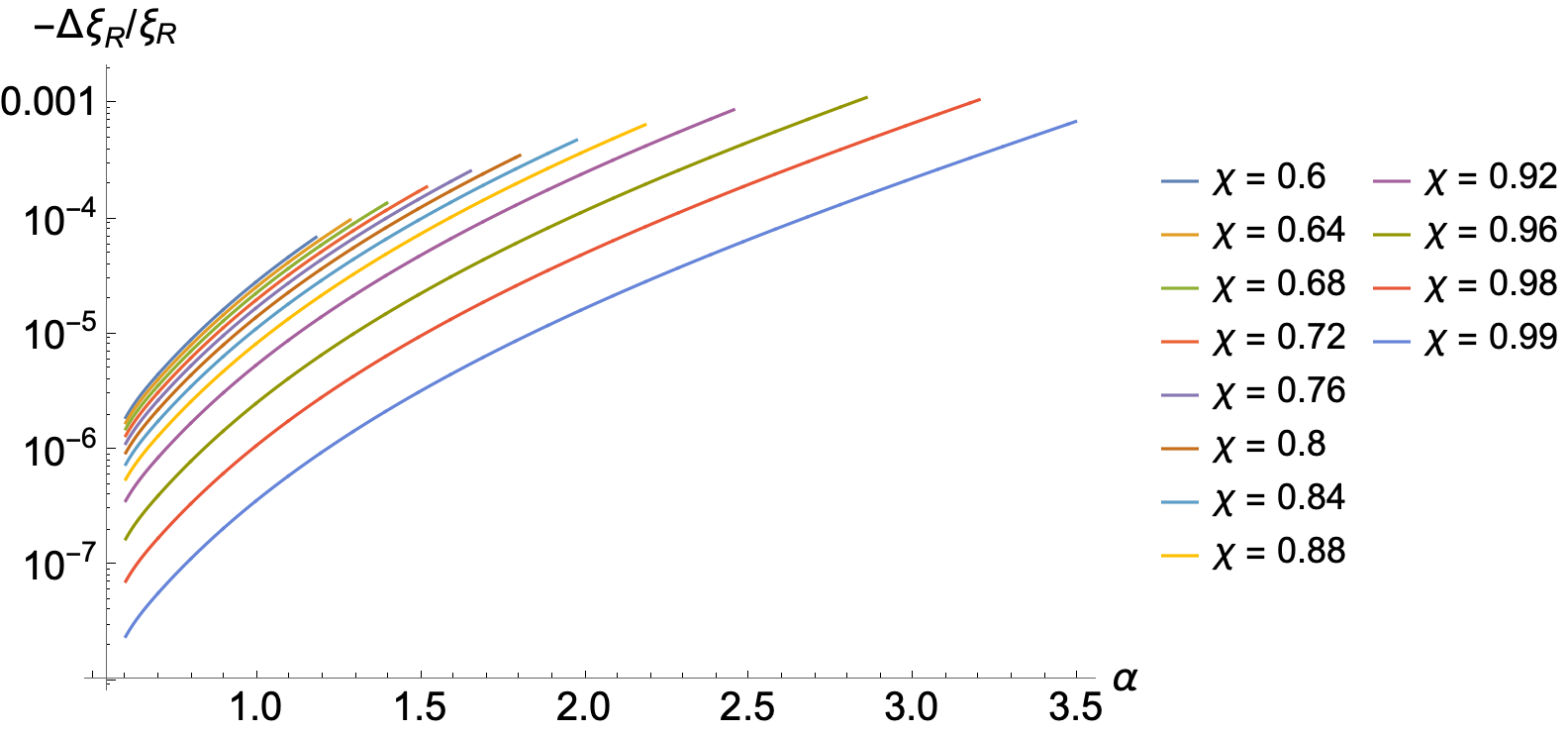}
    \caption{The fractional deviations between Fig$.$ \ref{fig:xir81} and the $\xi_{R}$ curves for various other spins.} 
    \label{fig:xir82}
\end{figure*}

\begin{figure*}
	\includegraphics[width=1.0\textwidth]{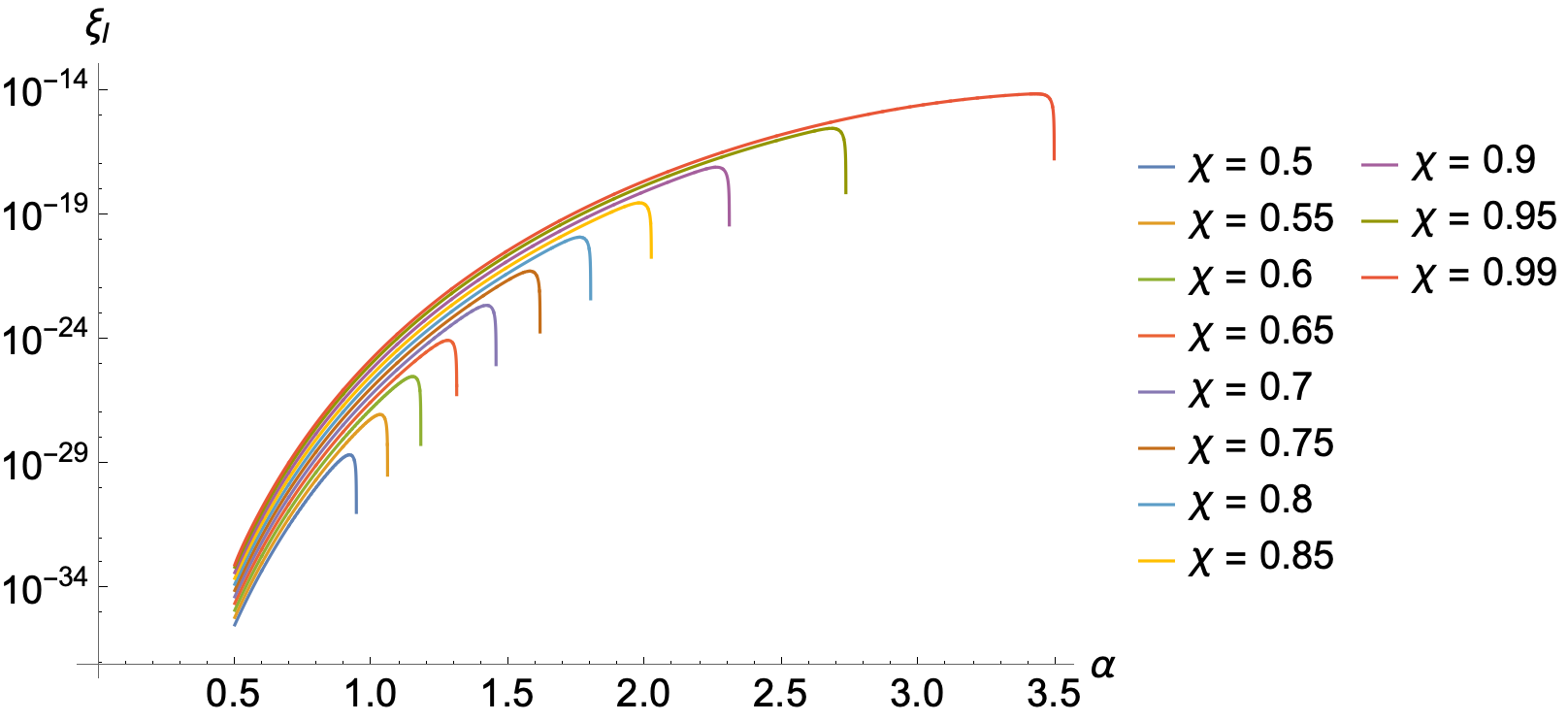}
	\caption{The imaginary part of the $n = 8$ bound-state eigenfrequency.}
	\label{fig:xii8}
\end{figure*}

The creation of an axion cloud corresponds to an instability of the Kerr space-time due to the presence of a massive scalar field. The amplifying mechanism, `superradiance', is the Penrose process in which rotational energy is extracted by a bosonic wave rather than by a particle. In the process, the Kerr BH loses mass and angular momentum, subject to the condition that its `irreducible mass' does not decrease. \\ \\
\indent In the Penrose scenario, a particle travelling through a BH's ergoregion can split in two, one of which falls into the hole, while the other escapes to infinity. If the orbital angular momentum of the infalling particle is of opposite sign to that of the hole, the BH's loses rotational energy to the escaping particle: Energy has been extracted from the ergoregion.  \\ \\ 
\indent The story for waves runs analogously: An incident wave with amplitude $\mathcal{I}$ splits into a part transmitted into the BH (with amplitude $\mathcal{T}$) and a part which escapes (the reflected wave with amplitude $\mathcal{R}$). If the transmitted wave is counter-rotating, the rotational energy of the BH decreases, leading to an outgoing wave with $\mathcal{R} > \mathcal{I}$. \\ \\
\indent The novelty of a massive scalar field is that its mass acts like a mirror: Unlike a massless field, a massive field can become trapped in a bound-orbit, leading to continuous extraction of rotational energy. The end result of the runaway amplification is a macroscopic scalar field bound-state -- the `axion cloud'. In an astrophysical context, rather than a wave incoming from infinity, the initial seed for superradiance can be any arbitrary quantum fluctuation in the scalar field, even if the field is in its classical ground state \cite{2019PhRvD..99h4042I}\cite{2020PhRvD.102f3020Z}$.$ As a result, the growth of an axion cloud begins immediately after the birth of a BH. \\ \\
\indent An axion cloud's binding energy (which determines the GW frequency) and growth timescale depend on the dynamics of the scalar field. For the scenario we have adopted, the field obeys the Klein-Gordon equation on the Kerr space-time. The Kerr metric describes an axisymmetric, neutral, and rotating black hole:
\begin{eqnarray}
    \label{kerr}
    ds^{2} = -\bigg[1 - \frac{2GMr}{c^{2}\rho^{2}} \bigg]c^{2}dt^{2} - \frac{4GMar \sin^{2}\theta}{c^{2}\rho^{2}}cdtd\phi \nonumber \\ + \frac{\rho^{2}}{\Delta}dr^{2} + \rho^{2}d\theta^{2} + \bigg[r^{2} + a^{2} + \frac{2GMa^{2}r\sin^{2}\theta}{c^{2}\rho^{2}} \bigg]\sin^{2}\theta d\phi^{2} \nonumber \\
\end{eqnarray}
where $M$ is the BH mass, $J$ is the BH angular momentum, $\rho^{2} \equiv r^{2} + a^{2}\cos^{2}\theta$, $a \equiv J/(Mc)$ is the Kerr parameter, and $\Delta \equiv r^{2} - 2r_{g}r + a^{2}$, where we have defined the gravitational radius $r_{g} \equiv GM/c^{2}$. In terms of the dimensionless Kerr parameter, $\chi \equiv a/r_{g} = Jc/(GM^{2})$, the inner and outer horizons -- the two roots of $\Delta = (r - r_{+})(r - r_{-})$ -- are
\begin{equation}
    r_{\pm} = r_{g}\Big[1 \pm \sqrt{1 - \chi^{2}} \Big]
\end{equation}
It follows that $\chi$ is restricted to the interval 
\begin{equation}
    0 < \chi < 1
\end{equation}
\indent The event horizon is located at $r = r_{+}$, and the angular velocity of the horizon is
\begin{equation}
    \Omega_{H} = \frac{c \chi}{2 r_{+}}
\end{equation}
\indent The scalar field obeys the Klein-Gordon equation,
\begin{equation}
    \Big[\nabla_{\mu}\nabla^{\mu} - m_{*}^{2} \Big]\Phi(\Vec{x}, t) = 0
    \label{kleingordon2}
\end{equation}
where $\nabla_{\mu}$ is the covariant derivative with respect to the Kerr metric, and, as mentioned in the text, $m_{*}$ has the quantum-mechanical interpretation as the reciprocal of the boson's Compton wavelength. In Boyer-Lindquist coordinates, the Klein-Gordon equation is separable via the ansatz
\begin{equation}
    \Phi(\Vec{x}, t) = \text{Re}\Big[ e^{-i \omega t}e^{i m \phi} S(\theta) R(r)\Big]
\label{kgansatz2}
\end{equation}
\indent Invoking the identity
\begin{equation}
    \nabla_{\mu}\nabla^{\mu}\Phi = \frac{1}{\sqrt{-g}}\partial_{\mu}\Big[\sqrt{-g} g^{\mu \nu} \partial_{\nu}\Phi \Big]
\end{equation}
$$\sqrt{-g} = \rho^{2}\sin \theta$$
the Klein-Gordon equation separates into two $2^{\text{nd}}$-order linear homogeneous ODE's for $R(r)$ and $S(\theta)$:
\begin{equation}
    \label{eqn:angular}
    \mathcal{D}_{\theta}[S] + \bigg[\chi^{2}\alpha^{2} \big(\xi^{2} - 1 \big)\cos^{2}\theta -\frac{m^{2}}{\sin^{2}\theta} + \Lambda \bigg]S(\theta) = 0
\end{equation}
\begin{eqnarray}
    \label{eqn:radial}
    \mathcal{D}_{r}[R] + \Bigg[\alpha^{2}\xi^{2}\big(r^{2} + \chi^{2}\big)^{2} - 4\chi m \alpha \xi r +  m^{2}\chi^{2} \nonumber \\ - \Delta\Bigg(\alpha^{2}r^{2} + \chi^{2}\alpha^{2}\xi^{2} + \Lambda \Bigg) \Bigg]R(r) = 0 \nonumber \\
\end{eqnarray}
\begin{equation}
    \mathcal{D}_{\theta} \equiv \frac{1}{\sin \theta}\frac{d}{d\theta}\bigg[\sin \theta \frac{d}{d\theta} \bigg], \ \ \ \  \mathcal{D}_{r} \equiv \Delta\frac{d}{dr}\bigg[\Delta\frac{d}{dr}\bigg]
\end{equation} \\
\indent We have expressed the decoupled equations in terms of the dimensionless variables ($\chi,\ \alpha$ and $\xi$) used in the main text$.$ The radial coordinate in \ref{eqn:radial} is measured in units of $r_{g}$. \\ \\
\indent Bound-state solutions must go to zero at infinity and be in-going at the event horizon$.$ The in-going condition means that $R(r) \propto e^{-ikr_{*}}$ as $r_{*} \rightarrow -\infty $, with $r_{*}$ the Kerr tortoise coordinate which maps the event horizon to $-\infty$,
\begin{equation}
    \frac{dr_{*}}{dr} = \frac{r^{2} + a^{2}}{\Delta}
\end{equation} \\
\indent This means that plane waves at the event horizon ($r_{*} \rightarrow -\infty$) can only move `to the left', i.e$.$ into the black hole. \\ \\ 
\indent The spectra of both bound-states and BH quasi-normal modes  can be found via Leaver's continued-fraction method \cite{1985RSPSA.402..285L} \cite{2007PhRvD..76h4001D}$.$ The radial function $R(r)$ is represented by an infinite series,
\begin{equation}
    R(r) = (r - r_{+})^{-i\sigma}(r - r_{-})^{i\sigma + \beta - 1}e^{qr}\sum_{n = 0}^{\infty}a_{n}\bigg(\frac{r - r_{+}}{r - r_{-}} \bigg)^{n}
\end{equation}
\begin{equation}
    \sigma = \frac{\alpha\big(1 + \sqrt{1 - \chi^{2}}\big)\big(\xi - \xi_{\text{crit}}\big)}{\sqrt{1 - \chi^{2}}}
\end{equation}
\begin{equation}
    q = \alpha\sqrt{1 - \xi^{2}}
\end{equation}
\begin{equation}
    \beta = \frac{\alpha^{2}\big(1 - 2\xi^{2} \big)}{q}
\end{equation}
(The quantity we denote by $\beta$ is the same as the quantity denoted by $\chi$ in Ref$.$ \cite{2007PhRvD..76h4001D}.) With this ansatz, \ref{eqn:radial} implies a three-term recurrence relation for the unknown coefficients $a_{n}$,
\begin{equation}
    \begin{cases}
        \alpha_{n}a_{n + 1} + \beta_{n}a_{n} + \gamma_{n}a_{n - 1} = 0, \ \ \ \ n = 1, 2, \ldots \\
        a_{1} = -\frac{\beta_{0}}{\alpha_{0}}a_{0}
    \end{cases}
\end{equation}
where the coefficients $\alpha_{n}$, $\beta_{n}$ and $\gamma_{n}$ are defined by
\begin{equation}
    \begin{cases}
        \alpha_{n} = n^{2} + (c_{0} + 1)n + c_{0} \\
        \beta_{n} = -2n^{2} + (c_{1} + 2) + c_{3} \\
        \gamma_{n} = n^{2} + (c_{2} - 3)n + c_{4}
    \end{cases}
\end{equation}
and $c_{0}, c_{1}, c_{2}, c_{3}$ and $c_{4}$ are given by
\begin{equation}
    c_{0} = 1 - 2 i \alpha  \xi -\frac{2 i \left(\alpha  \xi -\frac{m \chi }{2}\right)}{\sqrt{1-\chi ^2}} \ \ \ \ \ \ \ \ \ \ \ \ \ \ \ \ \ \ \ \ \ \ \ \ \ \ \ \
\end{equation}
\begin{eqnarray}
    \nonumber c_{1} = -4 + 4i\Big[\alpha  \xi - i\alpha\sqrt{1-\xi ^2} \left(1 + \sqrt{1 - \chi^2}\right)\Big] \ \ \ \ \ \ \ \ \ \ \\ 
    \nonumber + \frac{4i \left(\alpha \xi -\frac{m \chi}{2}\right)}{\sqrt{1 - \chi^2}} - \frac{2 \Big[\alpha ^2 \xi ^2+\alpha ^2 \left(1-\xi^2\right)\Big]}{\alpha \sqrt{1 - \xi^2}} \ \ \ \ \ \ \ \\
\end{eqnarray}
\begin{eqnarray}
    \nonumber c_{2} = 3 - 2i \alpha \xi - \frac{2 \Big[\alpha^2 \left(1 - \xi^2\right)-\alpha^2 \xi^2\Big]}{\alpha \sqrt{1-\xi^2}} - \frac{2 i \left(\alpha \xi -\frac{m \chi }{2}\right)}{\sqrt{1 - \chi^2}} \\
\end{eqnarray}
\begin{equation}
    \begin{split}
         c_{3} & = \frac{2i \left(\alpha \xi - i\alpha  \sqrt{1 -\xi^2}\right)^3}{\alpha \sqrt{1 - \xi^2}} + \chi^2\alpha^{2}\left(1 - \xi^2\right) \\
         & - \Lambda_{lm} - 1 +\ 2\sqrt{1 - \chi^2}\left(\alpha \xi - i\alpha \sqrt{1 - \xi^2}\right)^2 \\
         & + 2im\chi\alpha \sqrt{1 - \xi^2} - \frac{\left(\alpha \xi - i\alpha \sqrt{1 - \xi^2}\right)^2}{\alpha \sqrt{1 - \xi^2}} \\
         & + 2\alpha\sqrt{1 - \xi^2}\sqrt{1 -\chi^2} \\ 
         & + \frac{2i}{\sqrt{1 - \chi^2}}
    \Bigg[1 + \frac{\left(\alpha\xi - i\alpha \sqrt{1 - \xi^2}\right)^2}{\alpha 
   \sqrt{1 - \xi^2}}\Bigg]\bigg[\alpha\xi - \frac{m\chi}{2}\bigg]
    \end{split}
\end{equation}
\begin{eqnarray}
    \nonumber c_{4} = \frac{\left(\alpha \xi - i\alpha \sqrt{1 - \xi^2}\right)^4}{\alpha^2 \left(1 - \xi^2\right)} + \frac{2i\xi \left(\alpha \xi - i\alpha\sqrt{1 - \xi^2}\right)^2}{\sqrt{1 - \xi^2}} \\ 
    - \frac{2i \left(\alpha \xi - i\alpha \sqrt{1 - \xi^2}\right)^2 \left(\alpha \xi
   -\frac{m\chi}{2}\right)}{\alpha \sqrt{1 - \xi^2}\sqrt{1 - \chi^2}}  \ \ \ \ \ \ \ \ \ \ \ \
\end{eqnarray} \\
\indent The series coefficients are related by an infinite continued-fraction \cite{1967SIAMR...9...24G}
\begin{equation}
    \frac{a_{n + 1}}{a_{n}} = -\frac{\gamma_{n + 1}}{\beta_{n + 1} - \frac{\alpha_{n + 1}\gamma_{n + 2}}{\beta_{n + 2} - \ldots}}
\end{equation}
\indent Continued-fractions are commonly written in the slightly-less cumbersome notation
\begin{equation}
    \frac{a_{n + 1}}{a_{n}} = -\frac{\gamma_{n + 1}}{\beta_{n + 1} -}\frac{\alpha_{n + 1}\gamma_{n + 2}}{\beta_{n + 2}-}\frac{\alpha_{n + 2}\gamma_{n + 3}}{\beta_{n + 3}-}\ldots
\end{equation}\\
\indent Since $a_{1}/a_{0} = -\beta_{0}/\alpha_{0}$, we obtain a condition whose roots are the desired bound-state frequencies:
\begin{equation}
    \label{eqn:contfracradial}
    \beta_{0} - \frac{\alpha_{0}\gamma_{1}}{\beta_{1} -}\frac{\alpha_{1}\gamma_{2}}{\beta_{2} -}\frac{\alpha_{2}\gamma_{3}}{\beta_{3} -}\ldots = 0
\end{equation} \\
\indent Strictly speaking, the radial and angular eigenvalues, $\xi$ and $\Lambda$, must be found simultaneously. Leaver's method can also be applied to \ref{eqn:angular} \cite{2006PhRvD..73b4013B}, resulting in a continued-fraction condition analogous to \ref{eqn:contfracradial}. We then have two equations for the two unknowns. \\ \\
\indent Conveniently, we can reduce the root-finding problem to merely solving \ref{eqn:contfracradial} by using the $Mathematica$ function `SpheroidalEigenvalue'. With the change of variable $z = \cos \theta$, and in terms of the following quantities,
\begin{equation}
    \gamma^{2} \equiv \chi^{2}\alpha^{2}(1 - \xi^{2})
\end{equation}
\begin{equation}
    \lambda \equiv \Lambda - \gamma^{2}
\end{equation}
the angular equation \ref{eqn:angular} takes the standard form implemented in $Mathematica$:
\begin{equation}
    (1 - z^{2})\frac{d^{2}S}{dz^{2}} - 2z\frac{dS}{dz} + \Big[\gamma^{2}(1 - z^{2}) + \lambda - \frac{m^{2}}{1 - z^{2}} \Big]S(z) = 0
\end{equation} \\
\indent `SpheroidalEigenvalue' yields $\lambda$, and `SpheroidalPS' yields $S(z)$. The continued-fraction equation \ref{eqn:contfracradial}, with $\Lambda$ replaced by $\gamma^{2}\ +$ `SpheroidalEigenvalue', can then be solved for $\xi$ with the $Mathematica$ function \textbf{FindRoot}.  \\ \\
\indent For our axion cloud simulations, we have needed to compute $\xi$ for bound-states up to, and including, $n=9$. As an example, we have plotted the real \& imaginary parts of the $n=8$ bound-state in Figs$.$ \ref{fig:xir81}, \ref{fig:xir82} and \ref{fig:xii8}; Figs$.$ \ref{fig:xir81} and \ref{fig:xir82} are analogous to Figs$.$ \ref{fig:xir1} and \ref{fig:xir2}.


\begin{figure*}
	\includegraphics[width=0.7\textwidth]{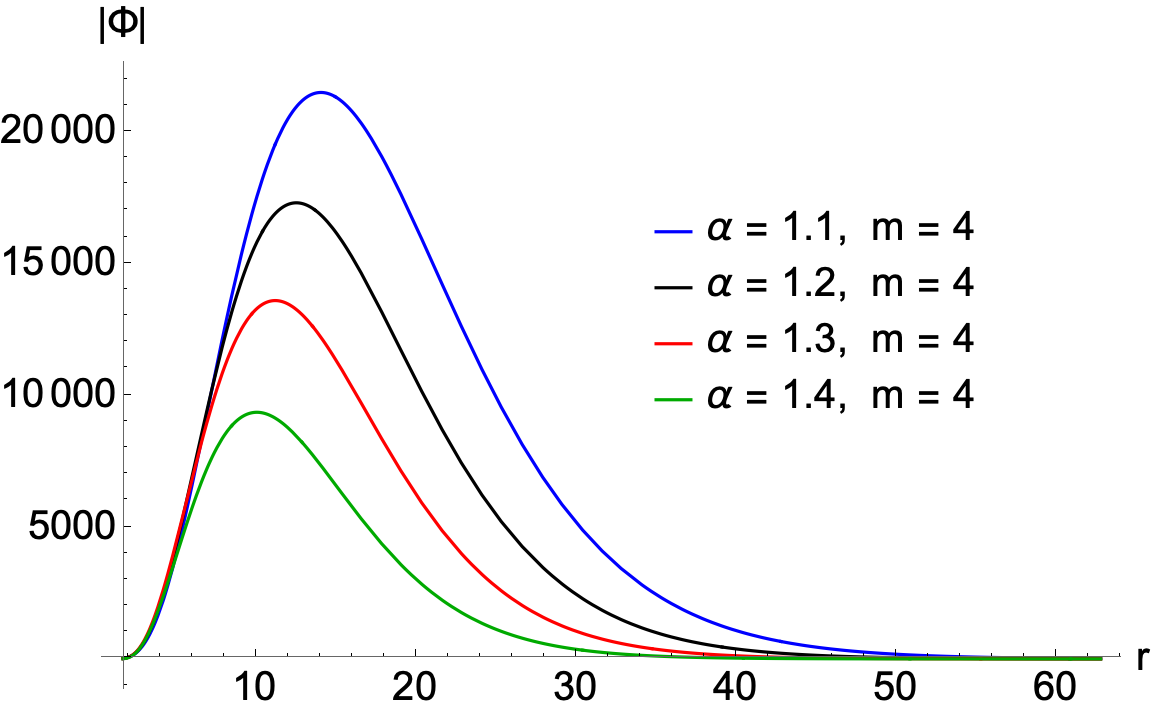}
	\caption{The radial profiles $\big|\Phi(r; t=0, \theta = \pi/2, \phi = 0)\big|$ of the $n = 5$, $l=m=4$ superradiant bound-state for a BH of spin $\chi = 0.99$ at four consecutive values of $\alpha$. The peak of the axion cloud shifts towards the BH with increasing $\alpha$. This makes sense by comparison with the hydrogen atom: For the $l = n - 1$ states of the hydrogen atom, the most probable radius $r_{\text{mp}}$ is inversely proportional to the electromagnetic fine-structure constant $\alpha_{\text{EM}}$: $r_{\text{mp}} = n^{2}a_{\text{B}} \propto n^{2}/\alpha_{\text{EM}}$, where $a_{\text{B}}$ is the Bohr radius. While $\alpha_{\text{EM}}$ is actually a constant, the analog for scalar field bound-states in Kerr, $\alpha \propto \mu M$, is different for each BH. As a result, for fixed $n$, the most probable radius for the $l = n - 1$ bound-states decreases with $\alpha$.}
	\label{fig:radialprofiles1}
\end{figure*}
\begin{figure*}
	\includegraphics[width=0.7\textwidth]{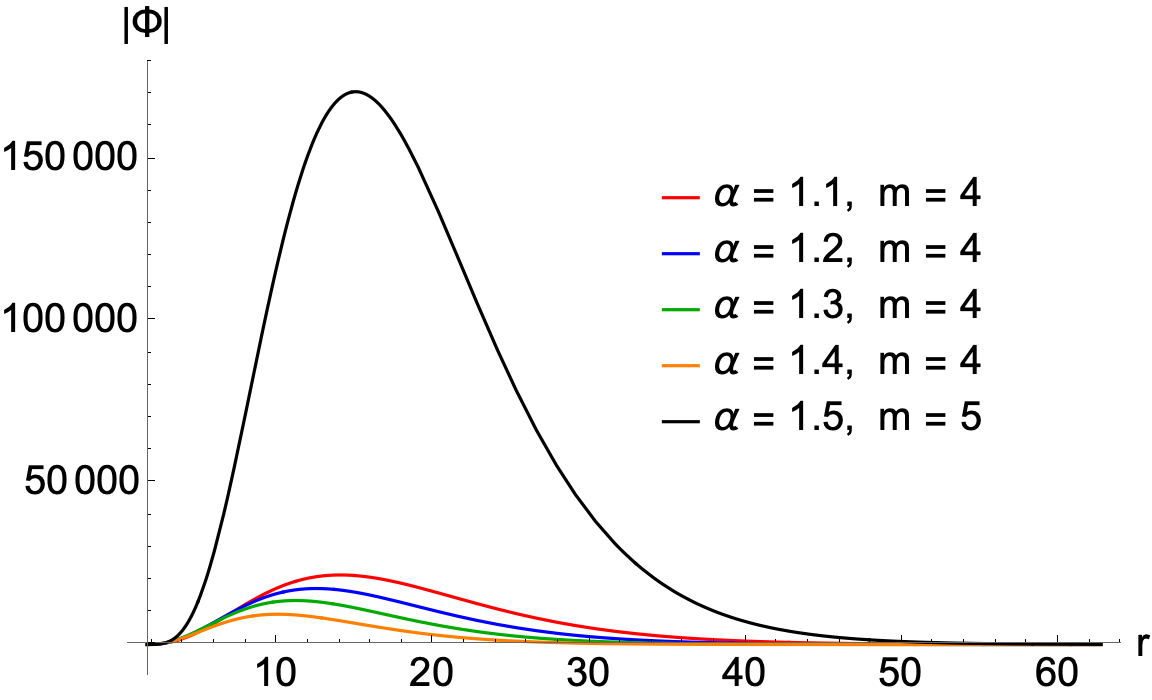}
	\caption{The same four radial profiles of Fig$.$ \ref{fig:radialprofiles1}, with the addition of the lowest superradiant bound-state for $\alpha = 1.5$ (with the BH spin still $\chi = 0.99$). This bound-state is $m = 5$, and its peak is to the right of the $m = 4$ states. This is consistent with the hydrogen-atom expectation that $r_{\text{mp}}$ increases with $n$.}
	\label{fig:radialprofiles2}
\end{figure*}
